\documentclass[]{report}   % list options between brackets

\usepackage{bm}
\usepackage[spanish,english]{babel}
\usepackage{amsfonts}
\usepackage{amssymb}
\usepackage{graphicx}
\usepackage[latin1]{inputenc}
\usepackage{hyperref}
\newcommand{\LL}{\mathcal{L}}

\newcommand{\M}{\mathcal{M}}
\newcommand{\U}{\mathcal{U}}
\newcommand{\Hcal}{\mathcal{H}}

\begin{document}

\title{Introduction\footnote{Chapter written for the book ``Open Questions in Cosmology", Edited by: Gonzalo J. Olmo, InTech Publishing, Rijeka, Croatia, (2012), ISBN 978-953-51-0880-1. The properly formatted version (28 pages) can be freely downloaded from the {publisher's website}: http://www.intechopen.com/books/open-questions-in-cosmology} to Palatini theories of gravity and nonsingular cosmologies. }   % type title between braces
\author{Gonzalo J. Olmo\\ {\small Departamento de F\'{i}sica Te\'{o}rica \& IFIC }\\ {\small Universidad de Valencia - CSIC} \\ {\small Burjassot 46100, Valencia,} \\ {\small Spain}}         % type author(s) between braces
\date{December, 2012}    % type date between braces

\maketitle

\begin{abstract}
 These notes are a summary of lectures given at the {\it Instituto de Astronom\'{i}a} of the {\it Universidad Nacional Aut\'{o}noma de M\'{e}xico} (UNAM), the {\it Dipartimento di Fisica} of the {\it Universit\`{a} degli studi di Salerno} (Italy), and the {\it Instituto de Física Corpuscular} of the {\it Universitat de Val\`{e}ncia} (Spain) during the year 2012. \\
Basic mathematical aspects of Palatini theories of gravity, which are constructed assuming that metric and connection are independent geometrical entities, are briefly introduced and discussed, and followed by a detailed derivation of the field equations of some rather general Lagrangian theories. Applications to the early universe are considered paying special attention to the avoidance of the big bang singularity. The analysis of non-singular cosmologies carried out in arXiv:$1005.4136$ [gr-qc]  is extended to include several perfect fluids. 
\end{abstract}

\addtocounter {chapter} {1}
\tableofcontents

\section{Introduction}

The impact of Einstein's fundamental idea of gravitation as a curved space-time phenomenon on our current understanding of the Universe has been enormously successful. A key aspect of his celebrated theory of General Relativity (GR) is that 
 the spatial sections of four dimensional space-time need not be Euclidean. The Minkowskian description is just an approximation  valid on (relatively) local portions of space-time. On larger scales, however, one must consider deformations induced by the matter on the geometry, which must be dictated by some set of field equations. In this respect, the predictions of GR are in agreement with experiments in scales that range from millimeters to astronomical units, scales in which weak and strong field phenomena can be observed [\cite{Will93}]. The theory is so successful in those regimes and scales that it is generally accepted that it should work also at larger and shorter scales, and at weaker and stronger regimes. The validity of these assumptions, obviously, is not guaranteed {\it a priori} regardless of how beautiful and elegant the theory might appear. Therefore, not only must we  keep confronting the predictions of the theory with experiments and/or  observations at new scales, but also we have to demand  theoretical consistency with the other physical interactions and, in particular, in the quantum regime. \\

For the above reasons, we believe that scrutinizing the implicit assumptions and mathematical structures behind the classical formulation of GR could help better understand the starting point of some current approaches that go beyond our standard model of gravitational physics. At the same time, this could provide new insights useful to 
address from a different perspective some current open questions, such as the existence of black hole and big bang singularities or the cosmic speedup problem.  In this sense, Einstein himself stated that ''{\it the question whether the structure of [the spacetime] continuum is Euclidean, or in accordance with Riemann's general scheme, or otherwise, is \ldots  a physical question which must be answered by experience, and not a question of a mere convention to be selected on practical grounds}'' [\cite{Feigl-Brodbeck}].  
From these words it follows that questioning the regime of applicability of the Riemannian nature of the geometry associated with the gravitational field and considering more general frameworks are legitimate questions that should be explored by all available means (theoretical and experimental). These are some of the basic points to be addressed in this work.\\ 

In this chapter we explore in some detail the implications of relaxing the Riemannian condition on the geometry by allowing the connection to be determined from first principles, not by choice or convention. This approach, known as metric-affine or Palatini formalism [\cite{MyReview}], assumes that metric and connection are equally fundamental and independent geometrical entities. In consequence, any geometrical theory of gravity formulated in this approach must provide enough equations to determine the form of the metric and the connection (within the unavoidable indeterminacy imposed by the underlying gauge freedom). We derive and discuss the field equations of a rather general family of Palatini theories and then focus on two particular subfamilies which have attracted special attention in recent years, namely, $f(R)$ and $f(R,Q)$ theories. The interest in studying these particular theories lies in their ability to avoid (or soften in some cases) big bang and black hole singularities and their relation with recent approaches to quantum gravity. Here we will focus on the early-time cosmology of such theories. 

The content is organized as follows. We begin by briefly reviewing in section  \ref{sec:Basics}  the basics of differentiable manifolds with affine and metric structures, to emphasize that metric and connection are equally fundamental and independent geometrical objects. In section \ref{sec:FieldEqs} a derivation of the field equations for a generic action depending on the metric and the Riemann tensor is presented taking into account also the presence of torsion. In section \ref{sec:PerFluids} we discuss a particular family of Lagrangians of the form $f(R,R_{\mu\nu}R^{\mu\nu})$ in combination with perfect fluid matter, and prepare the notation and field equations needed to study the dynamics of those theories. We then focus on the early-time characteristics of isotropic and anisotropic homogeneous cosmologies \ref{sec:cosmo} and show that nonsingular  bouncing solutions exist for $f(R)$ and $f(R,Q)$ models (subsections \ref{sec:f(R)} and \ref{sec:f(R,Q)}, respectively). We conclude with a discussion of the results presented and point out some open questions that should be addressed in the future.  

\section{Differentiable manifolds, affine connections, and the metric. \label{sec:Basics}}

In this section we quickly review some of the mathematical structures needed to construct a geometric theory of the gravitational interactions. The goal is to put forward  that metric and connection are equally fundamental and independent geometrical entities, an aspect usually overlooked in the construction of phenomenological extensions of GR. We will thus be more sketchy than mathematically accurate. For a more exhaustive and precise discussion of these topics see your favorite book on differentiable manifolds (or, for instance,  [\cite{Nakahara}]). \\ 

In the geometric description of gravitational theories, one begins by identifying physical events with points on an n-dimensional manifold $\M$. The next natural step is to provide this manifold with a differentiable structure. One then labels the points $p\in \M$ with a  set of charts $(\U_i,\varphi_i)$, where the $\U_i$ are subsets of $\M$ and $\varphi_i$ are maps from $\U_i$ to $\mathbb{R}^n$ (or an open subset of $\mathbb{R}^n$) such that every $p\in \M$ lies in at least one of the charts $(\U_i,\varphi_i)$. If for any two charts $(\U_i,\varphi_i)$ and $(\U_j,\varphi_j)$ that overlap at some nonzero subset of points the map $\varphi_i \circ {\varphi_j}^{-1}$ is not just continuous but differentiable, then we say that $\M$ is a differentiable manifold. \\
Since the Euclidean view of vectors as arrows connecting two points of the manifold is not valid in general, to get a consistent definition we need to introduce first the concept of curve and tangent vector to a curve at a point.  We thus say that a smooth curve $\gamma(t)$ in $\M$ is a differentiable map that to each point of a segment associates a point in $\M$, $\gamma(t):$ $t\in[0,1]\to \M$. In a chart $(\U,\varphi)$, the points of the curve have the following coordinate representation: $x=\varphi(p_t)=\varphi\circ \gamma(t)$. If we consider now a function $f$ on $\M$, where $f$ is a map that to every $p\in \M$ assigns a real number ($f: \M \to \mathbb{R}$), the rate of change of $f$ along the curve $\gamma(t)$ using the coordinates of the chart $(\U,\varphi)$ is given by 
\begin{equation}
\frac{df(\varphi\circ\gamma(t))}{dt}=\frac{df(x(t))}{dt}=\frac{\partial f}{\partial x^\mu}\frac{dx^\mu(t)}{dt}\equiv X^\mu(t)\frac{\partial f}{\partial x^\mu} \ ,
\end{equation} 
where we have defined the components of the tangent vector to the curve in this chart as $X^\mu\equiv dx^\mu(t)/dt$. Vectors can thus be seen as differential operators $X=X^\mu \partial_\mu$ whose action on functions is of the form $X[f]=X^\mu \partial_\mu f$, thus providing a natural notion of directional derivative for functions. The set $\{e_\mu\equiv \partial_\mu\}$ defines a (coordinate) basis of the tangent space of vectors at the point p, which we denote $T_p\M$. Obviously, vectors exist without specifying the coordinates. Under changes of coordinates, we have $V=V^\mu e_\mu= \tilde{V}^\alpha \tilde{e}_\alpha=\tilde{V}^\alpha \frac{\partial x^\mu}{\partial \tilde{x}^\alpha}e_\mu$, which implies the well-known transformation law $V^\mu=\tilde{V}^\alpha \frac{\partial x^\mu}{\partial \tilde{x}^\alpha}$ for  the vector components. \\
When a vector is assigned smoothly to each point of $\M$, it is called a vector field over $\M$. Each component of a vector field is thus a smooth function from $\M\to \mathbb{R}$. 
 Given a vector field $X$, an integral curve of $X$ is defined as the curve whose tangent vector coincides with $X$. For infinitesimal displacements of magnitude $\epsilon$ in the direction of $X$, a given point $p$ of coordinates $x^\mu$ 
 becomes $\sigma_\epsilon^\mu(x)=x^\mu+\epsilon X^\mu(x)$. This transformation also induces a correspondence between vectors of the tangent spaces $T_x\M$ and $T_{\sigma_\epsilon(x)}\M$. 
The effect of these transformations on a vector field $Y(x)$ leads to the concept of Lie derivative, whose action on vector fields is defined as 
\begin{equation}
\LL_X Y=  \left[ X^{\nu}\partial_{\nu} Y^{\mu}(x)-Y^{\nu}\partial_{\nu} X^{\mu}(x)\right] e_{\mu}\equiv [X,Y] \ . 
\end{equation}
This derivative operator is independent of the choice of coordinates and follows naturally from the differential structure of the manifold. It satisfies a number of useful properties such as bilinearity in its two arguments, $\LL_X (Y+Z)= \LL_X Y+ \LL_X Z$,  $\LL_{X+Y} Z= \LL_{X}Z+ \LL_Y Z$, and the chain rule $\LL_X fY=(\LL_X f)  Y+f\LL_X Y$, with $\LL_X f=X[f]$. Though the Lie derivative provides a natural {\it directional derivative} for functions, it does not work in the same way for vectors and tensors of higher rank. In fact, since the partial derivatives of the vector $X$ appear explicitly in $\LL_X Y$, two vectors whose components at a given point have the same values but whose partial derivatives at the point differ do not yield a vector that points in the same direction, i.e., they are not proportional. Therefore, in order to introduce a proper notion of {\it directional derivative} for vectors and tensors, we need to introduce a new structure called {\bf connection} which specifies how vectors (and tensors in general) are transported along a curve. \\

{\bf Manifolds with a connection.} We are thus going to introduce a derivative operator, which we denote by $\nabla$, such that given two vector fields $X$ and $Y$ we obtain a new vector field $Z$ defined by $Z\equiv \nabla_X Y$. This derivative operator must be bilinear in its two arguments, $\nabla_X (Y+Z)=\nabla_X Y+\nabla_X Z$, $\nabla_{X+Y}Z=\nabla_{X}Z+\nabla_{Y}Z$, must satisfy the chain rule $\nabla_X (fY)=(\nabla_X f) Y+f\nabla_X Y$, with $\nabla_X f=X[f]$, and must also behave as a natural directional derivative in the sense that $\nabla_{fX}Y=f\nabla_X Y$ to guarantee that any two proportional vectors yield a result that points in the same direction. In a given coordinate basis, we have $\nabla_X Y=X^\mu \nabla_{e_\mu} (Y^\nu e_{\nu})=X^\mu \left( e_\mu[Y^\nu]e_\nu+Y^\nu\nabla_{e_\mu}e_\nu\right)$. If our manifold is $m-$dimensional, defining $m^3$ functions called connection coefficients $\Gamma^\lambda_{\mu\nu}$ by $\nabla_{e_\mu} e_\nu\equiv \Gamma_{\mu\nu}^\lambda e_\lambda$ we find that the last requirement, $\nabla_{fX}Y=f\nabla_X Y$, is naturally satisfied. We thus find  that 
\begin{equation}
\nabla_X Y=X^\mu\left[\frac{\partial Y^\lambda}{\partial x^\mu}+\Gamma^\lambda_{\mu\nu}Y^\nu\right]e_\lambda \ .
\end{equation}
The connection coefficients specify how the basis vectors change from point to point and, in principle, can be arbitrarily defined. Under changes of coordinates, these coefficients transform as follows:
\begin{equation}
\nabla_{e_\mu}e_\nu\equiv\Gamma^\lambda_{\mu\nu}e_\lambda=\frac{\partial \tilde{x}^\alpha}{\partial x^\mu}\nabla_{\tilde{e}_\alpha} \left(\frac{\partial \tilde{x}^\beta}{\partial x^\nu}\tilde{e}_\beta\right)=\frac{\partial \tilde{x}^\alpha}{\partial x^\mu}\left[\frac{\partial {x}^\lambda}{\partial \tilde{x}^\nu}\frac{\partial^2 \tilde{x}^\gamma}{\partial x^\lambda \partial x^\nu}+\frac{\partial \tilde{x}^\beta}{\partial x^\nu}\tilde{\Gamma}^\gamma_{\alpha\beta}\right]\tilde{e}_\gamma=
\Gamma^\lambda_{\mu\nu}\frac{\partial\tilde{x}^\gamma}{\partial x^\lambda}\tilde{e}_\gamma \ ,
\end{equation}
which implies 
\begin{equation}
\Gamma^\lambda_{\mu\nu}=\frac{\partial {x}^\lambda}{\partial \tilde{x}^\gamma}\frac{\partial \tilde{x}^\alpha}{\partial x^\mu}\frac{\partial \tilde{x}^\beta}{\partial x^\nu}\tilde{\Gamma}^\gamma_{\alpha\beta}+
\frac{\partial {x}^\lambda}{\partial \tilde{x}^\gamma}\frac{\partial^2 \tilde{x}^\gamma}{\partial {x}^\mu\partial x^\nu} \ .
\end{equation}
This transformation law indicates that the connection coefficients do not transform as tensorial quantities. Therefore, the connection cannot have an intrinsic geometrical meaning as a measure of how much a manifold is curved. As intrinsic geometric objects, we can define the torsion tensor
\begin{equation}
T(X,Y)=\nabla_X Y-\nabla_Y X-[X,Y] \ ,
\end{equation}
and the Riemann curvature tensor 
\begin{equation}
R(X,Y)Z=\nabla_X \nabla_Y Z-\nabla_Y\nabla_X Z-\nabla_{[X,Y]}Z \ .
\end{equation}
In a coordinate basis, these tensors have the following components:
\begin{eqnarray}\label{eq:torsion}
T(e_\mu,e_\nu)&=&\left(\Gamma^\lambda_{\mu\nu}-\Gamma^\lambda_{\nu\mu}\right)e_\lambda \ , \\
R(e_\mu,e_\nu)e_\lambda &=& \left[\partial_\mu \Gamma^\beta_{\nu\lambda}-\partial_\nu \Gamma^\beta_{\mu\lambda}+
\Gamma^\kappa_{\nu\lambda}\Gamma^\gamma_{\mu\kappa}-\Gamma^\kappa_{\mu\lambda}\Gamma^\gamma_{\nu\kappa}\right]e_\gamma \ .
\end{eqnarray}

With the introduction of the connection, one can define the notion of parallel transport. Given a curve $\gamma(t)$ such that its tangent vector in a given chart has coordinates $X^\mu=dx^\mu(t)/dt$, we say that a vector $Y$ is parallel transported along $\gamma(t)$ if $\nabla_X Y=0$. In components, this equation reads $\frac{dY^\mu}{dt}+\Gamma^\mu_{\alpha\beta}\frac{dx^\alpha(t)}{dt} Y^\beta=0$, where $d/dt\equiv X^\mu\partial_\mu$. Geodesics are defined as those curves which are parallel transported along themselves, namely, $\nabla_X X=0$ or $\frac{d X^\mu}{dt}+\Gamma^\mu_{\alpha\beta}X^\alpha X^{\beta}=0$.

{\bf Manifolds with a metric.} So far we have been able to construct a number of geometrical objects such as directional derivatives of vectors and tensors in general, the torsion and Riemann tensors, geodesic curves, \ldots without the need to introduce a metric tensor. A metric tensor provides a notion of distance between nearby points and allows, among other things, to determine lengths, angles, areas, and volumes of objects which are locally defined in space-time. Formally, a (pseudo-Riemannian) metric tensor is a  symmetric bilinear form that at each $p\in \M$ satisfies $g_p(U,V)=g_p(V,U)$  for any two vectors $U,V\in T_p\M$ and $g_p(U,V)= 0$ for any $U\in T_p\M$ iff $V=0$. The metric tensor allows to define an inner product between vectors and also gives rise to an isomorphism between $T_p\M$ and the dual space of one-forms $T_p^*\M$.  In a coordinate basis, it can be represented by $g=g_{\mu\nu}dx^{\mu}\bigotimes dx^{\nu}$, where the differentials ${dx^\mu}$ form a basis of $T_p^*\M$. In manifolds with a metric, one can impose a particular relation between the metric and the connection by demanding that the scalar product of any two vectors which are parallel transported along any curve remains covariantly constant. This condition can be translated into\footnote{From now on we use the more standard notation $\nabla_\mu\equiv \nabla_{e_\mu}$} $\nabla_\mu g_{\alpha\beta}=0$, which implies that (recall that $T^\rho_{\beta\sigma}\equiv \Gamma^\rho_{\beta\sigma}-\Gamma^\rho_{\sigma\beta}$)
\begin{equation}
2\Gamma^\lambda_{(\mu\nu)}+\left({T^\rho_{\nu\sigma}}g_{\rho\mu}+{T^\rho_{\mu\sigma}}g_{\rho\nu}\right)g^{\sigma\lambda}=
g^{\lambda\rho}\left[\partial_\mu g_{\rho\nu}+\partial_\nu g_{\rho\mu}-\partial_\rho g_{\mu\nu}\right] \ .
\end{equation}
From the right-hand side of this equation, one defines the Levi-Civita connection as 
\begin{equation}
L^\lambda_{\mu\nu}\equiv \frac{g^{\lambda\rho}}{2}\left[\partial_\mu g_{\rho\nu}+\partial_\nu g_{\rho\mu}-\partial_\rho g_{\mu\nu}\right] \ .
\end{equation}
From this definition it follows that when the torsion vanishes, the connection is symmetric and coincides with the Levi-Civita connection. In that case, when $\Gamma^\lambda_{\mu\nu}=L^\lambda_{\mu\nu}$, we say that the associated geometry is Riemannian.  It should be noted that though  connections are not tensors, the difference between any two connections is a tensor. This, in particular, allowed us to construct the torsion tensor. With more generality, when the manifold is provided with a metric,  any connection $\Gamma^\lambda_{\mu\nu}$ can be expressed as 
\begin{equation}\label{eq:G-G}
\Gamma^\lambda_{\mu\nu}=L^\lambda_{\mu\nu}+A^\lambda_{\mu\nu}  \ ,
\end{equation}
where $A^\lambda_{\mu\nu} $ is a tensor (which needs not be symmetric in its lower indices). Therefore, Palatini theories of gravity, in which metric and connection are regarded as independent fields, can be seen as theories in which an additional rank-three tensor field $A^\lambda_{\mu\nu} $ has been added to the gravitational Lagrangian.

\section{Dynamics of Palatini theories \label{sec:FieldEqs}}

From the above quick review of the properties of  differentiable manifolds with metric and affine structures, it is clear that  metric and connection are equally fundamental and independent geometrical entities. In the construction of theories of gravity based on geometry, we will thus assume this independence and will require those theories to yield equations that allow to determine both the metric and the connection and the possible relations between them. For simplicity, we will assume that the matter is only coupled to the metric (which is consistent with the experimental tests of the equivalence principle [\cite{Will93}]) but will allow an independent connection to appear in the gravitational sector of the theory. As pointed out above, this is equivalent to having, besides the metric, a rank-three gravitational tensor field. From a geometric perspective, this possibility seems much more  natural and fundamental than considering, for instance, scalar fields in the gravitational sector, though scalar-tensor theories have traditionally received much more attention in the literature. \\

We begin by deriving the field equations of Palatini theories in a very general case and then consider some simplifications to make contact with the literature. For a generic Palatini theory in which the connection appears through the Riemann tensor or contractions of it, the action can be written as follows [\cite{OlmoERE11}]
\begin{equation}\label{eq:f-action}
S=\frac{1}{2\kappa^2}\int d^4x \sqrt{-g}f(g_{\mu\nu},{R^\alpha}_{\beta\mu\nu})+S_m[g_{\mu\nu},\psi] \ ,
\end{equation}
where $S_m$ is the matter action, $\psi$ represents collectively the matter fields, $\kappa^2$ is a constant with suitable dimensions (if $f=R$, then $\kappa^2=8\pi G$),  and
\begin{equation}
 {R^\alpha}_{\beta\mu\nu}=\partial_\mu\Gamma_{\nu\beta}^\alpha-\partial_\nu\Gamma_{\mu\beta}^\alpha+\Gamma_{\mu\lambda}^\alpha\Gamma_{\nu\beta}^\lambda-\Gamma_{\nu\lambda}^\alpha\Gamma_{\mu\beta}^\lambda
\end{equation}
 represents the components of the Riemann tensor, the field strength of the connection $\Gamma^\alpha_{\mu\beta}$. Note that since the connection is determined dynamically, i.e., we assume independence between the metric and affine structures of the theory, we cannot assume any {\it a priori} symmetry in its lower indices. This means that in the variation of the action to obtain the field equations we must bear in mind that $\Gamma^\alpha_{\beta\gamma}\neq  \Gamma^\alpha_{\gamma\beta}$, i.e., we admit the possibility of nonvanishing torsion. It should be noted that in GR energy and momentum are the sources of curvature, while torsion is sourced by the spin of particles [\cite{Kibble-Sciama}]. 
The fact that torsion is usually not considered in introductory courses on gravitation may be rooted in the educational tradition of this subject and the fact that the spin of particles was discovered many years after the original formulation of GR by Einstein. Another reason may be that the effects of torsion are very weak in general, except at very high densities, where the role of torsion becomes dominant and may even avoid the formation of singularities (see [\cite{Poplawski:2010kb}] for a recent discussion and earlier literature on the topic). For these reasons, and to motivate and facilitate the exploration of the effects of torsion in extensions of GR, our derivation of the field equations will be as general as possible (within reasonable limits). \\
 We will assume a symmetric metric tensor $g_{\mu\nu}=g_{\nu\mu}$ and the usual definitions for the Ricci tensor $R_{\mu\nu}\equiv{R^\rho}_{\mu\rho\nu}$ and the Ricci scalar $R\equiv g^{\mu\nu}R_{\mu\nu}$. The variation of the action (\ref{eq:f-action}) with respect to the metric and the connection can be expressed as
\begin{eqnarray}\label{eq:var1-f}
\delta S&=&\frac{1}{2\kappa^2}\int d^4x \sqrt{-g}\left[\left(\frac{\partial f}{\partial g^{\mu\nu}} -\frac{f}{2}g_{\mu\nu} \right)\delta g^{\mu\nu} + \frac{\partial f}{\partial {R^\alpha}_{\beta\mu\nu}} \delta {R^\alpha}_{\beta\mu\nu}\right]+\delta S_m \ .
\end{eqnarray}
Straightforward manipulations show that $\delta {R^\alpha}_{\beta\mu\nu}$ can  be written as
\begin{equation}
\delta {R^\alpha}_{\beta\mu\nu}= \nabla_\mu \left(\delta \Gamma^\alpha_{\nu\beta}\right)-\nabla_\nu \left(\delta\Gamma^\alpha_{\mu\beta}\right)+2S^\lambda_{\mu\nu}\delta\Gamma^\alpha_{\lambda\beta} \ ,
\end{equation}
 where $S^\lambda_{\mu\nu}\equiv ( \Gamma^\lambda_{\mu\nu}-\Gamma^\lambda_{\nu\mu})/2$ now represents the torsion tensor tensor (note the additional $1/2$ factor as compared to our initial definition in \ref{eq:torsion}). From now on we will use the notation ${P_\alpha}^{\beta\mu\nu}\equiv \frac{\partial f}{\partial {R^\alpha}_{\beta\mu\nu}}$.  In order to put the $\delta {R^\alpha}_{\beta\mu\nu}$ term in (\ref{eq:var1-f}) in suitable form, we need to note that 
\begin{equation}\label{eq:step1}
I_\Gamma=\int d^4x \sqrt{-g} {P_\alpha}^{\beta\mu\nu}\nabla_\mu \delta \Gamma^\alpha_{\nu\beta}=\int d^4x \left[\nabla_\mu(\sqrt{-g}J^\mu)-\delta \Gamma^\alpha_{\nu\beta}\nabla_\mu\left(\sqrt{-g} {P_\alpha}^{\beta\mu\nu}\right)\right] \ ,
\end{equation}
where $J^\mu\equiv {P_\alpha}^{\beta\mu\nu}\delta \Gamma^\alpha_{\nu\beta}$. Since, in general, $\nabla_\mu(\sqrt{-g}J^\mu)=\partial_\mu(\sqrt{-g}J^\mu)+2S^\sigma_{\sigma \mu}\sqrt{-g}J^\mu$, we find that (\ref{eq:step1}) can be written as 
\begin{equation}\label{eq:step2}
I_\Gamma=\int d^4x \left[\partial_\mu(\sqrt{-g}J^\mu)-\delta \Gamma^\alpha_{\nu\beta}\left\{\nabla_\mu\left(\sqrt{-g} {P_\alpha}^{\beta\mu\nu}\right)-2S^\sigma_{\sigma \mu}\sqrt{-g}{P_\alpha}^{\beta\mu\nu}\right\}\right] \ .
\end{equation}
Using this result, (\ref{eq:var1-f}) becomes
\begin{eqnarray}\label{eq:var2-f}
\delta S&=&\frac{1}{2\kappa^2}\int d^4x \left[\sqrt{-g}\left(\frac{\partial f}{\partial g^{\mu\nu}} -\frac{f}{2}g_{\mu\nu} \right)\delta g^{\mu\nu}+\partial_\mu\left(\sqrt{-g}J^\mu\right) \right. \\ 
&+& \left.\left\{-\frac{1}{\sqrt{-g}}\nabla_\mu \left(\sqrt{-g}{P_\alpha}^{\beta[\mu\nu]} \right)+S^\nu_{\sigma\rho}{P_\alpha}^{\beta\sigma\rho}+2S^\sigma_{\sigma\mu}{P_\alpha}^{\beta[\mu\nu]}\right\}2\sqrt{-g}\delta \Gamma^\alpha_{\nu\beta}\right]+\delta S_m \ . \nonumber
\end{eqnarray}
We thus find that the field equations can be written as follows
\begin{eqnarray}\label{eq:gmn}
\kappa^2 T_{\mu\nu}&=&\frac{\partial f}{\partial g^{(\mu\nu)}} -\frac{f}{2}g_{\mu\nu}  \\
\kappa^2{H_\alpha}^{\nu\beta}&=&-\frac{1}{\sqrt{-g}}\nabla_\mu \left(\sqrt{-g}{P_\alpha}^{\beta[\mu\nu]} \right)+S^\nu_{\sigma\rho}{P_\alpha}^{\beta\sigma\rho}+2S^\sigma_{\sigma\mu}{P_\alpha}^{\beta[\mu\nu]} \ , \label{eq:Gamn}
\end{eqnarray}
where ${P_\alpha}^{\beta[\mu\nu]}=({P_\alpha}^{\beta\mu\nu}-{P_\alpha}^{\beta\nu\mu})/2$, $T_{\mu\nu}=-\frac{2}{\sqrt{-g}}\frac{\delta S_m}{\delta g^{\mu\nu}}$ is the energy-momentum tensor of the matter, and $ {H_\alpha}^{\nu\beta}=-\frac{1}{\sqrt{-g}}\frac{\delta S_m}{\delta \Gamma^\alpha_{\nu\beta}}$ represents the coupling of matter to the connection. For simplicity,   from now on we will assume that  ${H_\alpha}^{\nu\beta}=0$.  Eq. (\ref{eq:Gamn}) can be put in a more convenient form if the connection is decomposed into its symmetric and antisymmetric (torsion) parts, $\Gamma^\alpha_{\mu\nu}=C^\alpha_{\mu\nu}+S^\alpha_{\mu\nu} $, such that $\nabla_\mu A_\nu=\partial_\mu A_\nu-C^\alpha_{\mu\nu} A_\alpha-S^\alpha_{\mu\nu} A_\alpha=\nabla_\mu^C A_\nu-S^\alpha_{\mu\nu} A_\alpha$ and $\nabla_\mu \sqrt{-g}=\nabla_\mu^C \sqrt{-g}-S^\alpha_{\mu\alpha}\sqrt{-g}$. By doing this, (\ref{eq:Gamn}) turns into 
\begin{equation}
\kappa^2{H_\alpha}^{\nu\beta}=-\frac{1}{\sqrt{-g}}\nabla_\mu^C \left(\sqrt{-g}{P_\alpha}^{\beta[\mu\nu]} \right)+S^\lambda_{\mu\alpha}{P_\lambda}^{\beta[\mu\nu]}-S^\beta_{\mu\lambda}{P_\alpha}^{\lambda[\mu\nu]} \ . \label{eq:Gamn2}
\end{equation}

\subsection{Example: f(R,Q) theories}
Eqs. (\ref{eq:gmn}) and (\ref{eq:Gamn2}) can be used to write the field equations for the metric and the connection for specific choices of the Lagrangian $f(g_{\mu\nu},{R^\alpha}_{\beta\mu\nu})$. To make contact with the literature [\cite{Olmo:2011sw}], [\cite{BO2010}] ,[\cite{OSAT}], we now focus on the case $f(R,Q)=f(g^{\mu\nu}R_{\mu\nu},g^{\mu\nu}g^{\alpha\beta}R_{\mu\alpha}R_{\nu\beta})$. 
For this family of Lagrangians, we obtain
\begin{equation}
{P_\alpha}^{\beta\mu\nu}={\delta_\alpha}^\mu M^{\beta\nu}={\delta_\alpha}^\mu\left(f_R g^{\beta\nu}+2f_Q R^{\beta\nu}\right) \ ,
\end{equation}
where $f_X=\partial_X f$. Inserting this expression in (\ref{eq:Gamn2}) and tracing over $\alpha$ and $\nu$, we find that $\nabla_\lambda^C[\sqrt{-g}M^{\beta\lambda}]=(2\sqrt{-g}/3)[S^\sigma_{\lambda\sigma}M^{\beta\lambda}+(3/2)
S^\beta_{\lambda\mu}M^{\lambda\mu}]$. Using this result, the connection equation can be put as follows
\begin{equation}
\frac{1}{\sqrt{-g}}\nabla_\alpha^C\left[\sqrt{-g}M^{\beta\nu}\right]=S^\nu_{\alpha\lambda}M^{\beta\lambda}-S^\nu_{\beta\lambda}M^{\lambda\nu}-S^\lambda_{\alpha\lambda}M^{\beta\nu}+\frac{2}{3}\delta_\alpha^\nu S^\sigma_{\lambda\sigma}M^{\beta\lambda}
\end{equation}
The symmetric and antisymmetric combinations of this equation lead, respectively, to 
\begin{eqnarray}\label{eq:symm}
\frac{1}{\sqrt{-g}}\nabla_\alpha^C\left[\sqrt{-g}M^{(\beta\nu)}\right]=S^\nu_{\alpha\lambda}M^{[\beta\lambda]}-S^\beta_{\alpha\lambda}M^{[\nu\lambda]}-S^\lambda_{\alpha\lambda}M^{(\beta\nu)}+\frac{S^\sigma_{\lambda\sigma}}{3}\left(\delta_\alpha^\nu M^{\beta\lambda}+\delta_\alpha^\beta M^{\nu\lambda}\right)\\ 
\frac{1}{\sqrt{-g}}\nabla_\alpha^C\left[\sqrt{-g}M^{[\beta\nu]}\right]=S^\nu_{\alpha\lambda}M^{(\beta\lambda)}-S^\beta_{\alpha\lambda}M^{(\nu\lambda)}-S^\lambda_{\alpha\lambda}M^{[\beta\nu]}+\frac{S^\sigma_{\lambda\sigma}}{3}\left(\delta_\alpha^\nu M^{\beta\lambda}-\delta_\alpha^\beta M^{\nu\lambda}\right)  \ . \label{eq:asymm}
\end{eqnarray}
Important simplifications can be achieved considering the new variables 
\begin{equation}\label{eq:newG}
\tilde{\Gamma}^\lambda_{\mu\nu}=\Gamma^\lambda_{\mu\nu}+\alpha \delta^\lambda_\nu S^\sigma_{\sigma\mu} \ ,
\end{equation}
and taking the parameter $\alpha=2/3$, which implies that $\tilde{S}^\lambda_{\mu\nu}\equiv \tilde{\Gamma}^\lambda_{[\mu\nu]}$ is such that $\tilde{S}^\sigma_{\sigma\nu}=0$. The symmetric and antisymmetric parts of $\tilde{\Gamma}^\lambda_{\mu\nu}$ are related to those of  ${\Gamma}^\lambda_{\mu\nu}$ by
\begin{eqnarray}\label{eq:newC}
\tilde{C}^\lambda_{\mu\nu}&=&C^\lambda_{\mu\nu}+\frac{1}{3}\left(\delta^\lambda_\nu S^\sigma_{\sigma\mu}+\delta^\lambda_\mu S^\sigma_{\sigma\nu}\right) \\
\tilde{S}^\lambda_{\mu\nu}&=&S^\lambda_{\mu\nu}+\frac{1}{3}\left(\delta^\lambda_\nu S^\sigma_{\sigma\mu}-\delta^\lambda_\mu S^\sigma_{\sigma\nu}\right) \label{eq:newS}
\end{eqnarray}
Using these variables, Eqs. (\ref{eq:symm}) and  (\ref{eq:asymm}) take the following compact form
\begin{eqnarray}\label{eq:symm2}
\frac{1}{\sqrt{-g}}\nabla_\alpha^{\tilde{C}}\left[\sqrt{-g}M^{(\beta\nu)}\right] &=&\left[\tilde{S}^\nu_{\alpha\lambda}g^{\beta\kappa}+
\tilde{S}^\beta_{\alpha\lambda}g^{\nu\kappa}\right]g^{\lambda\rho}M_{[\kappa\rho]} \\  \label{eq:asymm2}
\frac{1}{\sqrt{-g}}\nabla_\alpha^{\tilde{C}}\left[\sqrt{-g}M^{[\beta\nu]}\right] &=& \left[\tilde{S}^\nu_{\alpha\lambda}g^{\beta\kappa}-
\tilde{S}^\beta_{\alpha\lambda}g^{\nu\kappa}\right]g^{\lambda\rho}M_{(\kappa\rho)}  \ .
\end{eqnarray}  
In these equations, $M^{(\beta\nu)}=f_R g^{\beta\nu}+2f_Q R^{(\beta\nu)}(\Gamma)$, and $M^{[\beta\nu]}=2f_Q R^{[\beta\nu]}(\Gamma)$, where $R_{(\beta\nu)}(\Gamma)=R_{(\beta\nu)}(\tilde{\Gamma})$ and $R_{[\beta\nu]}(\Gamma)=R_{[\beta\nu]}(\tilde{\Gamma})+\frac{2}{3}\left(\partial_\beta S^\sigma_{\sigma\nu}-\partial_\nu S^\sigma_{\sigma\beta}\right)$. \\ 

In the recent literature on Palatini theories, only the torsionless case has been studied in detail. When torsion is considered in $f(R)$ theories, Eqs. (\ref{eq:symm2}) and (\ref{eq:asymm2}) recover the results presented in [\cite{MyReview}].  In general, those equations put forward that when the traceless torsion tensor $\tilde{S}^\nu_{\alpha\lambda}$ vanishes, the symmetric and antisymmetric parts of $M^{\beta\nu}$ decouple. The dynamics of these theories, therefore, can be studied in different levels of complexity. The simplest case  will be studied here and consists on setting ${S}^\nu_{\alpha\lambda}$ and $R_{[\mu\nu]}$ to zero. A more detailed discussion of the other cases can be found in [\cite{Olmo-Rubiera}].

\subsection{Volume-invariant and torsionless $f(R,Q)$ \label{sec:vol-inv}}

When the torsion is set to zero,  it can be shown [\cite{Schouten}], [\cite{Hehl:1994ue}] that the vanishing of $R_{[\mu\nu]}$ guarantees the existence of a volume element that is covariantly conserved by $\Gamma^\alpha_{\mu\nu}$. The rank-two tensor that defines that volume element must be a solution of (\ref{eq:symm2}), which in this case takes the form
\begin{equation}\label{eq:connection}
\nabla_\alpha^{\Gamma}\left[\sqrt{-g}\left(f_R g^{\beta\nu}+2f_Q R^{\beta\nu}(\Gamma)\right)\right]=0 \ . 
\end{equation}
Note that here $R^{\beta\nu}(\Gamma)$ is symmetric because we are taking $R_{[\mu\nu]}(\Gamma) =0$.  To obtain the solution of (\ref{eq:connection}),  we first consider (\ref{eq:gmn}) particularized to our theory (with ${S}^\nu_{\alpha\lambda}$ and $R_{[\mu\nu]}$ set to zero),
\begin{equation}
f_R R_{\mu\nu}-\frac{f}{2}g_{\mu\nu}+2f_QR_{\mu\alpha}{R^\alpha}_\nu = \kappa^2 T_{\mu\nu}\label{eq:met-varX} \ ,
\end{equation}
and rewrite it in the following form
\begin{equation}
f_R {B_\mu}^\nu-\frac{f}{2}{\delta_\mu}^\nu+2f_Q{B_\mu}^\alpha {B_\alpha}^\nu= \kappa^2 {T_\mu}^\nu\label{eq:met-varRQ1} \ ,
\end{equation}
where we have defined  ${B_\mu}^\nu\equiv R_{\mu\alpha}g^{\alpha\nu}$. This equation can be seen as a second-order algebraic equation for the matrix $\hat{B}$, whose components are ${[\hat{B}]_\mu}^\nu\equiv {B_\mu}^\nu$. The solutions to this equation imply that $\hat{B}$ is an algebraic function of the components of the stress-energy tensor ${T_\mu}^\nu$, i.e., $\hat{B}=\hat{B}(\hat{T})$. This relation is very important because it allows to express (\ref{eq:connection}) in the form 
\begin{equation}\label{eq:connection2}
\nabla_\alpha^{\Gamma}\left[\sqrt{-g}g^{\beta\lambda}\left(f_R \delta_\lambda^\nu +2f_Q {B_\lambda}^\nu\right)\right]=0 \ ,
\end{equation}
where now $f_R$, $f_Q$ and ${B_\alpha}^\nu$ are functions of the stress-energy tensor of the matter. The connection, therefore, can be obtained by elementary algebraic manipulations [\cite{OSAT}]. To do it, one defines a rank-two symmetric tensor $h^{\mu\nu}$ such that $\sqrt{-g}g^{\beta\lambda}\left(f_R \delta_\lambda^\nu +2f_Q {B_\lambda}^\nu\right)=\sqrt{-h}h^{\beta\nu}$, which turns (\ref{eq:connection2}) into the well-known equation $\nabla_\mu \left[\sqrt{-h}h^{\beta\nu}\right]=0$, and implies that $\Gamma^\alpha_{\beta\nu}$ is given by the Christoffel symbols of the tensor $h_{\mu\nu}$, i.e.,
\begin{equation}\label{eq:LC-h}
\Gamma^{\alpha}_{\beta\gamma}=\frac{h^{\alpha\rho}}{2}\left(\partial_\beta h_{\rho\gamma}+\partial_\gamma h_{\rho\beta}-\partial_\rho h_{\beta\gamma}\right) \ .
\end{equation}
From the defining expression of $h_{\mu\nu}$, one finds that the relation between $h_{\mu\nu}$ and $g_{\mu\nu}$ can be expressed as follows
\begin{equation} \label{eq:h-g}
h_{\mu\nu}=\sqrt{\det\hat\Sigma} {[\Sigma^{-1}]_\mu}^\alpha g_{\alpha\nu}  \  ,   \ h^{\mu\nu}=\frac{g^{\mu\alpha}{{\Sigma_\alpha} ^\nu}}{\sqrt{\det\hat\Sigma}} \ , 
\end{equation}
 where we have defined the matrix ${\Sigma_\alpha} ^\nu\equiv \left(f_R \delta_\alpha^\nu +2f_Q {B_\alpha}^\nu\right)$. With these relations and definitions, the field equations for the metric $h_{\mu\nu}$ can be written in compact form expressing (\ref{eq:met-varRQ1})  as ${B_\mu}^\alpha {\Sigma_\alpha}^\nu=\frac{f}{2}\delta_\mu^\nu+\kappa^2{T_\mu}^\nu$ and using the relation ${B_\mu}^\alpha {\Sigma_\alpha}^\nu=\sqrt{\det\hat\Sigma} R_{\mu\alpha}(h) h^{\alpha\nu}$ to obtain [\cite{Olmo:2011aw}]
\begin{equation}\label{eq:field-eqs}
{R_\mu}^\nu(h)=\frac{1}{\sqrt{\det\hat\Sigma}}\left(\frac{f}{2}\delta_\mu^\nu+\kappa^2{T_\mu}^\nu\right) \ .
\end{equation}
In general, it will be more convenient to work with the field equations for the auxiliary metric $h_{\mu\nu}$ because their form is more tractable. Nonetheless, if one insists on writing the field equations using the metric $g_{\mu\nu}$, one must note that the connection (\ref{eq:LC-h}) is related to the Levi-Civita connection of $g_{\mu\nu}$ by the tensor  [recall Eq.(\ref{eq:G-G})]
\begin{equation}
A^{\alpha}_{\beta\gamma}\equiv \Gamma^{\alpha}_{\beta\gamma}-L^{\alpha}_{\beta\gamma}= 
\frac{h^{\alpha\rho}}{2}\left[\nabla_\mu^{L}h_{\rho\nu}+\nabla_\nu^{L}h_{\rho\mu}-\nabla_\rho^{L}h_{\mu\nu}\right] \ .
\end{equation}
The Riemann tensors of $\Gamma^{\alpha}_{\beta\gamma}$ and $L^{\alpha}_{\beta\gamma}$ are thus related as follows 
\begin{equation}
{R^\alpha}_{\beta\mu\nu}(\Gamma)={R^\alpha}_{\beta\mu\nu}(L)+\nabla_\mu^L A^\alpha_{\nu\beta}-\nabla_\nu^L A^\alpha_{\mu\beta}+A^\lambda_{\nu\beta}A^\alpha_{\mu\lambda}-A^\lambda_{\mu\beta}A^\alpha_{\nu\lambda} \ ,
\end{equation}
which allows to express (\ref{eq:field-eqs}) in terms of the Ricci tensor of the metric $g_{\mu\nu}$, the usual covariant derivatives of  $L^{\alpha}_{\beta\gamma}$, and the matter. 

\section{ $f(R,Q)$ theories with a perfect fluid \label{sec:PerFluids}}

The explicit form of the matrix $\hat\Sigma$ that relates the metrics $h_{\mu\nu}$ and $g_{\mu\nu}$ can only be found once all the sources that make up $T_{\mu\nu}$ have been specified. In our discussion we will just consider a perfect fluid or a sum of non-interacting perfect fluids such that 
\begin{equation}\label{eq:Tmn}
T_{\mu\nu}=(\rho+P)u_\mu u_\nu+P g_{\mu\nu}
\end{equation}
with $\rho=\sum_i \rho_i$ and $P=\sum_i P_i$. In order to find an expression for $\hat\Sigma$, we first rewrite (\ref{eq:met-varRQ1}) using matrix notation as
\begin{equation}
2f_Q\hat{B}^2+f_R \hat{B}-\frac{f}{2}\hat{I} = \kappa^2 \hat{T} \label{eq:met-varRQ2} \ .
\end{equation}
Using (\ref{eq:Tmn}) this equation can be rewritten as follows
\begin{equation}
2f_Q\left(\hat B+\frac{f_R}{4f_Q}\hat I\right)^2=\left(\kappa^2 P+\frac{f}{2}+\frac{f_R^2}{8f_Q}\right)\hat I+\kappa^2(\rho+P)u_\mu u^\mu\label{eq:met-varRQ3} \ .
\end{equation} 
Denoting $\lambda^2\equiv \left(\kappa^2P+\frac{f}{2}+\frac{f_R^2}{8f_Q}\right)$ and making explicit the matrix representation, (\ref{eq:met-varRQ3}) becomes
\begin{equation}
2f_Q\left(\hat B+\frac{f_R}{4f_Q}\hat I\right)^2=\left(\begin{array}{lr}
 \lambda^2-\kappa^2(\rho+P) & \vec{0}  \\
\vec{0} & \lambda^2 \hat{I}_{3X3}
\end{array}\right)   \label{eq:met-varRQ4} \ ,
\end{equation} 
where $\hat{I}_{3X3}$ denotes 3-dimensional identity matrix. Since the right-hand side of (\ref{eq:met-varRQ4}) is a diagonal matrix, it is immediate to compute its square root, which leads to 
\begin{equation}
\sqrt{2f_Q}\left(\hat B+\frac{f_R}{4f_Q}\hat I\right)=\left(\begin{array}{lr}
 s_1\sqrt{\lambda^2-\kappa^2(\rho+P)} & \vec{0}  \\
\vec{0} & \lambda \hat{S}_{3X3}
\end{array} \right)  \label{eq:met-varRQ5} \ ,
\end{equation} 
where $s_1$ denotes a sign, which can be positive or negative, and $\hat{S}_{3X3}$ denotes a $3X3$ diagonal matrix with elements $\{s_i=\pm 1\}$. For consistency of the theory in the limit $f_Q\to 0$, we must have $s_1=1$ and $\hat{S}_{3X3}=\hat{I}_{3X3}$. This result allows to express $\hat \Sigma$ as follows
\begin{equation}
\hat \Sigma=\left(\begin{array}{lr}
 \sigma_1 & \vec{0}  \\
\vec{0} & \sigma_2 \hat{I}_{3X3}
\end{array} \right)  \label{eq:Sigma} \ ,
\end{equation} 
where $\sigma_1$ and $\sigma_2$ take the form
\begin{eqnarray}
\sigma_1&=&  \frac{f_R}{2}\pm \sqrt{2f_Q}\sqrt{\lambda^2-\kappa^2(\rho+P)}\nonumber\\
\sigma_2&=& \frac{f_R}{2}+\sqrt{2f_Q}\lambda \ . \label{eq:sigmas}
\end{eqnarray}
Note that we have kept the two signs $\pm$ in $\sigma_1$. The reason for this will be understood later, when particular models are considered. The point is that in some cases of physical interest, at high densities one should take the negative sign in front of the square root to guarantee that $\sigma_1$ is continuous and differentiable accross the point where the square root vanishes. This technical issue does not arise for $\sigma_2$.

\subsection{Workable models: $f(R,Q)=\tilde{f}(R)+\alpha Q$ \label{sec:model}}

So far we have made progress without specifying the form of the Lagrangian $f(R,Q)$. However, in order to find the explicit dependence of $R={B_\mu}^\mu$ and $Q={B_\mu}^\alpha{B_\alpha}^\mu$ with the $\rho$ and $P$ of the fluids, we must choose a Lagrangian explicitly. Restricting the function $f(R,Q)$ to the family 
$f(R,Q)=\tilde{f}(R)+\alpha Q$, we will see that it is possible to find the generic dependence of $Q$ with $\rho$ and $P$, while $R$ is found to depend only on the combination $T=-\rho+3P$ [\cite{OSAT}]. The reason for this follows from the trace of (\ref{eq:met-varX}) with $g^{\mu\nu}$, which for this family of Lagrangians gives the algebraic relation $R \tilde{f}_R-2\tilde{f}=\kappa^2T$ and implies that $R=R(T)$ (like in Palatini $f(R)$ theories). For these theories, we have that $f_Q=\alpha$, which is a constant. Therefore, from the trace of (\ref{eq:met-varRQ4}) we find
\begin{equation}
\sqrt{2f_Q}\left(R+\frac{f_R}{f_Q}\right)=
\sqrt{\lambda^2-\kappa^2(\rho+P)} +3\lambda\ ,
\end{equation}
which can be cast as
\begin{equation}
\left[\sqrt{2f_Q}\left(R+\frac{f_R}{f_Q}\right)-3\lambda\right]^2=
\lambda^2-\kappa^2(\rho+P)
\end{equation}
After a bit of algebra we find that 
\begin{equation}\label{eq:lambda1}
\lambda= \frac{\sqrt{2f_Q}}{8}\left[3\left(R+\frac{f_R}{f_Q}\right)\pm\sqrt{\left(R+\frac{f_R}{f_Q}\right)^2-\frac{4\kappa^2(\rho+P)}{f_Q}}\right]
\end{equation}
From this expression and the definition of $\lambda^2$,  we find
\begin{equation} \label{eq:Q}
\alpha Q=-\left(\tilde f+\frac{\tilde f_R^2}{4f_Q}+2\kappa^2P\right)+\frac{{f_Q}}{16}\left[3\left(R+\frac{\tilde f_R}{f_Q}\right)\pm\sqrt{\left(R+\frac{\tilde f_R}{f_Q}\right)^2-\frac{4\kappa^2(\rho+P)}{f_Q}}\right]^2 \ ,
\end{equation}
where $R$, $\tilde{f}$, and $\tilde f_R$ are functions of $T=-\rho+3P$.

\section{Nonsingular cosmologies in $f(R,Q)$ theories \label{sec:cosmo}}

The difficulties faced by GR to provide a consistent description of singularities and quantum phenomena at high energies (microscopic or Planck scales) is generally seen as an indication that we should go beyond the standard geometric structures to successfully quantize the theory and avoid singularities. This idea has motivated a variety of approaches that range from the consideration of higher-dimensional superstrings and other extended objects [\cite{strings}], to  non-commutative geometries  or   non-perturbative quantization methods [\cite{LQG1}], [\cite{LQG2}], [\cite{LQG3}] , to name just a few well-known cases. Unfortunately, the formidable task of building a satisfactory quantum theory of gravity is not yet complete. Moreover, even if we managed to get such a theory, we would still have to face the challenge of testing its predictions. In this sense, it should be noted that since the quantum gravitational regime is so far from our current and future experimental capabilities, our only hope might be to use the information available in the cosmic microwave background radiation to verify or rule out our theories [\cite{Agullo:2011xv}]. How much of the quantum gravitational regime could be contrasted with these yet-to-come theories is not clear. This is due, in part, because the theorized rapid accelerated expansion that took place during the inflationary period may have washed out many of the relevant proper signatures needed to distinguish the predictions of different quantum theories of gravity. \\

A conservative approach, therefore, consists on exploring the quantum properties and interactions of the matter fields in the very early universe using the well-established methods of quantum field theory in curved space-times [\cite{Parker-Toms}]. The success of this approach has been confirmed in combination with models of inflation and sheds relevant light on the mechanisms that may have caused the primordial spectra of scalar and tensorial perturbations [\cite{Inflation_books0}], [\cite{Inflation_books1}], [\cite{Inflation_books2}],[\cite{Inflation_books3}], [\cite{Inflation_books4}]. The applicability of this approach, however, becomes unreliable at increasing energies as the regime of the classical big bang singularity is approached and the quantum fluctuations of the gravitational field can no longer be neglected. At that stage, a complete quantum theory of gravity seems necessary to provide a consistent  description of the ongoing physical processes. Obviously, different quantum theories could lead to completely different quantum gravitational scenarios and, therefore, a generic quantum origin for the universe cannot be  guessed {\it a priori} by any logical means.  \\

In recent years, bouncing cosmological models have attracted much attention [\cite{Novello-2008}]. These are scenarios in which the big bang singularity is replaced by a quantum-induced bounce that connects an earlier phase of contraction with the subsequent expanding phase (in which we happen to exist). In such scenarios, aside from the quantum regime, the contracting and expanding phases are expected to asymptote an effective classical geometry whose dynamics, on consistency grounds, should match that of GR at low energies. In this context, and as an intermediate step between the quantum field theory approach in the (singular) curved background provided by GR and a (nonsingular) full theory of quantum gravity, one could consider the case of a smooth effective geometry free from big bang singularities on top of which quantum matter fields could still be treated perturbatively in a consistent way. This view would somehow disentangle the non-perturbative part of the quantum gravitational sector into an effective classical, nonsingular geometry, plus perturbative quantum corrections that propagate on top of the regular effective background. The absence of curvature singularities would make the treatment of quantum fields on the resulting geometry more reliable, and could help shed new light on the effects of the matter-gravity interaction in the very-early universe. \\

In the literature there exist many interesting examples of (quantum and non-quantum) cosmological models that avoid the big bang singularity by means of a bounce. Roughly, those models can be classified in two large groups, depending on whether they contain a modified gravitational sector or a modified matter sector (see \cite{Novello-2008} for details and a very complete list of references). Generically, modified gravity theories imply the existence of new dynamical degrees of freedom, such as gravitational scalar fields (like in scalar-tensor theories), higher-derivatives of the metric, extra dimensions, \ldots The consideration of exotic matter sources may be justified, in some cases,  from an effective field theory approach, such as in the case of non-linear theories of electrodynamics, which naturally arise in low-energy limits of string theories. \\
 In the remainder of this chapter, we are going to study bouncing cosmological models from the modified gravity perspective provided by the Palatini theories discussed above. This approach is particularly interesting because, despite being a modified-gravity approach,  the underlying mechanisms that modify the gravitational dynamics are not associated with new dynamical degrees of freedom or higher-derivative equations. In fact, it is the nontrivial role played by the matter in the determination of the space-time connection that induces nonlinearities in the matter sector that end up changing the dynamics at very high matter-energy densities. In this sense, it should be noted that the  gravitational field equations in vacuum exactly recover those of GR (with possibly a cosmological constant, depending on the particular Lagrangian chosen). For this reason, this type of theories can be regarded as a {\it minimal extension of the standard model of gravitational physics}, because they only appreciably depart from GR in regions that contain sources and when those sources reach the energy-density scales that characterize the correcting terms of the Lagrangian.   \\

\subsection{Homogeneous cosmologies in $f(R,Q)$ theories} \label{sec:Exp-Shear}

In this section we introduce the basic definitions and formulas needed to derive the equations for the evolution of the expansion and shear [\cite{Wald1984}] for an arbitrary Palatini $f(R,Q)$ theory of the kind presented in Section \ref{sec:vol-inv} .  These magnitudes will be very useful to extract information about the geometric properties of the space-time and to determine whether cosmic singularities are present or not.  We focus on homogeneous cosmologies of the Bianchi I type (a different expansion factor for each spatial direction) because that will allow us to test the rebustness of our results against deviations from the idealized Friedmann-Robertson-Walker space-times (same expansion rate in all the spatial directions). We will also particularize our results to the case of $f(R)$ theories, i.e., no dependence on $Q$. \\

We consider a Bianchi I spacetime with physical line element of the form
\begin{equation}
ds^2=g_{\mu\nu}dx^\mu dx^\nu=-dt^2+\sum_{i=1}^3 a_i^2(t)(dx^i)^2
\end{equation} 
In terms of this line element, using the relation between metrics (\ref{eq:h-g}) and the expression (\ref{eq:Sigma}) for the matrix $\hat\Sigma$ of a collection of perfect fluids, the nonzero components of the auxiliary metric $h_{\mu\nu}$ become
\begin{eqnarray}\label{eq:hmn}
h_{tt}&=& -\left(\frac{\sigma_2^2}{\sqrt{\sigma_1\sigma_2}}\right)\equiv -S \\
h_{ij}&=& \sqrt{\sigma_1\sigma_2} a_i^2 \delta_{ij}\equiv\Omega a_i^2 \delta_{ij} 
\end{eqnarray} 
The relevant Christoffel symbols associated with $h_{\mu\nu}$ are the following:
\begin{eqnarray}
\Gamma^t_{tt}&=& \frac{\dot S}{2S} \\
\Gamma^t_{ij}&=& \frac{\Omega a_i^2}{2S}\left[\frac{\dot\Omega}{\Omega}+\frac{2\dot a_i}{a_i}\right]\delta_{ij}\\
\Gamma^i_{tj}&=& \frac{\delta^i_j}{2}\left[\frac{\dot\Omega}{\Omega}+\frac{2\dot a_i}{a_i}\right]
\end{eqnarray}
The nonzero components of the corresponding Ricci tensor are
\begin{eqnarray}
R_{tt}(h)&=& -\sum_i\dot H_i-\sum_iH_i^2-\frac{3}{2}\frac{\ddot\Omega}{\Omega}+\frac{3}{4}\frac{\dot\Omega}{\Omega}\left(\frac{\dot S}{S}+\frac{\dot\Omega}{\Omega}\right)+\frac{1}{2}\left(\frac{\dot S}{S}-\frac{2\dot\Omega}{\Omega}\right)\sum_iH_i\\
R_{ij}(h)&=& \frac{\delta_{ij} a_i^2}{2}\frac{\Omega}{S}\left[2\dot H_i+\frac{\ddot\Omega}{\Omega}-\left(\frac{\dot\Omega}{\Omega}\right)^2+\frac{\dot\Omega}{\Omega}\sum_kH_k+\frac{1}{2}\frac{\dot\Omega}{\Omega}\left(\frac{3\dot\Omega}{\Omega}-\frac{\dot S}{S}\right)+\right.\nonumber \\ 
&+& \left.2H_i\left\{\sum_kH_k+\frac{1}{2}\left(\frac{3\dot\Omega}{\Omega}-\frac{\dot S}{S}\right)\right\}\right] \ ,
\end{eqnarray}
where $H_k\equiv \dot a_k/a_k$.  For completeness, we give an expression for the corresponding scalar curvature
\begin{equation}
R(h)= \frac{1}{S}\left[2\sum_k\dot H_k+\sum_k H_k^2+\left(\sum_k H_k\right)^2+\left(3\frac{\dot\Omega}{\Omega}-\left\{\frac{\dot S}{S}-\frac{\dot\Omega}{\Omega}\right\}\right)\sum_k H_k+3\frac{\ddot\Omega}{\Omega}-\frac{3}{2}\frac{\dot\Omega}{\Omega}\frac{\dot S}{S}\right]
\end{equation}

From the above formulas, one can readily find the corresponding ones in the isotropic, flat configuration by just replacing $H_i\to H$. For the spatially nonflat case, the $R_{tt}(h)$ component is the same as in the flat case. The $R_{ij}(h)$ component, however, picks up a new piece, $2K\gamma_{ij}$, where $\gamma_{ij}$ represents the nonflat spatial metric of $g_{ij}=a^2_i\gamma_{ij}$. The Ricci scalar then becomes $R(h)\to R^{K=0}(h)+\frac{6K}{a^2\Omega}$.\\ 

\subsection{Shear}

From the previous formulas and the field equation (\ref{eq:field-eqs}), we find that the combination ${R_i}^i-{R_j}^j$ (no summation over indices) leads to
\begin{equation}\label{eq:Rii-Rjj}
{R_i}^i-{R_j}^j=\frac{1}{S}\left[\dot H_{ij}+H_{ij}\left\{\sum_kH_k+\frac{1}{2}\left(\frac{3\dot\Omega}{\Omega}-\frac{\dot S}{S}\right)\right\}\right]=0 \ ,
\end{equation}
where we have defined $H_{ij}\equiv H_i-H_j$. Note that the final equality  ${R_i}^i-{R_j}^j=0$, follows from the fact that the right hand sides of ${R_i}^i$ and ${R_j}^j$ as given by (\ref{eq:field-eqs}) are equal. 
Expressing (\ref{eq:Rii-Rjj}) in the form 
\begin{equation} 
{R_i}^i-{R_j}^j=\frac{d}{dt}\left[\ln H_{ij}+\ln (a_1 a_2 a_3)+\ln \Omega^{3/2}-\ln S^{1/2} \right]=0 \ ,
\end{equation}
we see that it can be readily integrated regardless of the number and particular equations of state of the fluids involved. The result is 
\begin{equation}\label{eq:Hij}
H_{ij}=C_{ij}\frac{S^{\frac{1}{2}}}{\Omega^{\frac{3}{2}}}\frac{C_{ij}}{(a_1 a_2 a_3)}=\frac{C_{ij}}{\sigma_1}\frac{V_0}{V(t)} \ ,
\end{equation}
where the constants $C_{ij}=-C_{ji}$ satisfy the relation $C_{12}+C_{23}+C_{31}=0$, $V_0$ represents a reference volume, and $V(t)=V_0 a_1 a_2 a_3$ represents the volume of the universe.  It is worth noting that writing explicitly the three equations (\ref{eq:Hij}) and combining them in pairs, one can write the individual Hubble rates as follows
\begin{eqnarray}
H_1&=& \theta +\frac{\left(C_{12}-C_{31}\right)}{3\sigma_1} \left(\frac{V_0}{V(t)}\right)\nonumber \\
H_2&=& \theta +\frac{\left(C_{23}-C_{12}\right)}{3\sigma_1}\left(\frac{V_0}{V(t)}\right) \label{eq:Hi}\\
H_3&=& \theta +\frac{\left(C_{31}-C_{23}\right)}{3\sigma_1}\left(\frac{V_0}{V(t)}\right)\nonumber
\end{eqnarray}
where $\theta$  is the expansion of a congruence of comoving observers and is defined as $3\theta=\sum_i H_i$. Using these relations, the shear $\sigma^2=\sum_i\left(H_i-\theta\right)^2$ of the congruence takes the form
\begin{equation}\label{eq:shear}
\sigma^2=\frac{(C_{12}^2+C_{23}^2+C_{31}^2)}{9\sigma_1^2}\left(\frac{V_0}{V(t)}\right)^2 \ ,
\end{equation}
where we have used the relation $(C_{12}+C_{23}+C_{31})^2=0$.   

\subsection{Expansion}

We now derive an equation for the evolution of the expansion with time and a relation between expansion and shear.
From previous results, one finds that 
\begin{equation}\label{eq:Gtt}
G_{tt}(h)\equiv -\frac{1}{2}\sum_k H_k^2+\frac{1}{2}\left(\sum_k H_k\right)^2+\frac{\dot \Omega}{\Omega}\sum_k H_k+\frac{3}{4}\left(\frac{\dot \Omega}{\Omega}\right)^2
\end{equation}
In terms of the expansion and shear, this equation becomes 
\begin{equation}\label{eq:Gtt-1}
G_{tt}\equiv 3\left(\theta+\frac{\dot\Omega}{2\Omega}\right)^2-\frac{\sigma^2}{2} \ . 
\end{equation}
From the field equation (\ref{eq:field-eqs}), we find that 
\begin{equation}\label{eq:Gtt-rhs}
G_{tt}= \frac{f+\kappa^2(\rho+3P)}{2\sigma_1} \ ,
\end{equation}
which in combination with (\ref{eq:Gtt-1}) yields
\begin{equation}\label{eq:Hubble}
3\left(\theta+\frac{\dot\Omega}{2\Omega}\right)^2=\frac{f+\kappa^2(\rho+3P)}{2\sigma_1}+\frac{\sigma^2}{2} \ .
\end{equation}
For a set of non-interacting fluids with equations of state $w_i=P_i/\rho_i$, we have that $\Omega=\Omega(\rho_i, w_i)$ and, therefore, $\dot\Omega=\sum_i \Omega_{\rho_i}\dot\rho_i$, where $\Omega_{\rho_i}\equiv \partial\Omega/\partial {\rho_i}$. Since for those fluids the conservation equation is  $\dot\rho_i=-3\theta (1+\omega_i)\rho_i$, we find that $\dot\Omega=-3\theta \sum_i  (1+\omega_i) \rho_i \Omega_{\rho_i}$. With this result, (\ref{eq:Hubble}) can be written as 
\begin{equation}\label{eq:Hubble-pfluids}
3\theta^2 \left(1+\frac{3}{2}\Delta_1\right)^2=\frac{f+\kappa^2(\rho+3P)}{2\sigma_1}+\frac{\sigma^2}{2} \ ,
\end{equation}
where we have defined
\begin{equation}\label{eq:D1}
\Delta_1=-\sum_i(1+w_i)\rho_i\frac{\partial_{\rho_i}\Omega}{\Omega} \ .
\end{equation}
Note that in this last equation $w_i=w_i(\rho_i)$, i.e., they need not be constants.
For fluids with constant $w_i$, the conservation equation implies that their density depends on the volume of the universe according to $\rho_i(t)=\rho_i(t_0)\left(\frac{V_0}{V(t)}\right)^{1+w_i}$. This implies that once a particular Lagrangian is specified, the equations of state $P_i=w_i\rho_i$ are given, and the anisotropy constants $C_{ij}$ are chosen, the right-hand side of Eqs. (\ref{eq:shear}) and (\ref{eq:Hubble-pfluids}) can be parametrized in terms of $V(t)$. This, in turn, allows us to parametrize the $H_i$ functions of (\ref{eq:Hi}) in terms of $V(t)$ as well. This will be very useful later for our discussion of particular models. \\

In the isotropic case ($\sigma^2=0 \ , \theta=\dot{a}/a\equiv \Hcal$) with nonzero spatial curvature, (\ref{eq:Hubble-pfluids}) takes the following form:
\begin{equation}\label{eq:Hubble-iso}
\Hcal^2=\frac{1}{6\sigma_1}\frac{\left[f+\kappa^2(\rho+3P)-\frac{6K\sigma_2}{a^2}\right]}{\left[1+\frac{3}{2}\Delta_1\right]^2} 
\end{equation}

The evolution equation for the expansion can be obtained by noting that the $R_{ij}$ equations, which are of the form $R_{ij}\equiv(\Omega/2S)g_{ij}\left[\ldots\right]=(f/2+\kappa^2P)g_{ij}/\sigma_2$, can be summed up to give 
\begin{equation}
2(\dot{\theta}+3\theta^2)+\theta \left(\frac{6\dot{\Omega}}{\Omega}-\frac{\dot{S}}{S}\right)+\left\{\frac{\ddot{\Omega}}{\Omega}+\frac{1}{2}\frac{\dot{\Omega}}{\Omega}\left(\frac{\dot{\Omega}}{\Omega}-\frac{\dot{S}}{S}\right)\right\}=\frac{\left[f+2\kappa^2P\right]}{\sigma_1} \ .
\end{equation}

\subsection{Limit to $f(R)$ \label{sec:limit}}

We now consider the limit $f_Q\to 0$, namely, the case in which the Lagrangian only depends on the Ricci scalar $R$. Doing this we will obtain the corresponding equations for shear and expansion in the $f(R)$ case without the need for extra work. From the definitions of $\lambda^2$ (see below eq.(\ref{eq:met-varRQ3})), and  $\sigma_1$ and $\sigma_2$ in (\ref{eq:sigmas}), it is easy to see that in the limit $f_Q\to  0$ we get
\begin{eqnarray}
\sigma_1&\to&\sigma_2 \to f_R  \\
S &\to& \Omega \to f_R \ .
\end{eqnarray}
With these rules it is easy to see that $h_{\mu\nu}=f_R g_{\mu\nu}$, which makes (\ref{eq:field-eqs}) boil down to the expected field equations for Palatini $f(R)$ theories, namely, $f_R R_{\mu\nu}(h)-\frac{f}{2}g_{\mu\nu}=\kappa^2T_{\mu\nu}$.
Equation (\ref{eq:Hij}) turns into 
\begin{equation}\label{eq:Hij-f(R)}
H_{ij}=\frac{C_{ij}}{f_R}\frac{V_0}{V(t)}, 
\end{equation}
from which one can easily obtain expressions for $H_1$, $H_2$ and $H_3$ as in (\ref{eq:Hi}). 
The shear becomes
\begin{equation}\label{eq:shear-f(R)}
\sigma^2=\frac{(C_{12}^2+C_{23}^2+C_{31}^2)}{9f_R^2}\left(\frac{V_0}{V(t)}\right)^2  \ ,
\end{equation}
where $C_{12}+C_{23}+C_{31}=0$. The relation between expansion and shear for a collection of non-interacting perfect fluids now becomes
\begin{equation}\label{eq:Hubble-f(R)}
3{\theta^2}\left(1+\frac{3}{2}\tilde\Delta_1\right)^2=\frac{f+\kappa^2(\rho+3P)}{2f_R}+\frac{\sigma^2}{2}
\end{equation}
where $\tilde\Delta_1$ is given by (\ref{eq:D1}) but with $\Omega$ replaced by $f_R$. In the isotropic case with nonzero $K$ we find
\begin{equation}\label{eq:Hubble-iso-f(R)}
\Hcal^2=\frac{1}{6f_R}\frac{\left[f+\kappa^2(\rho+3P)-\frac{6K f_R}{a^2}\right]}{\left[1+\frac{3}{2}\tilde\Delta_1\right]^2}  \ .
\end{equation}

\subsection{Bouncing $f(R)$ Cosmologies  \label{sec:f(R)}}

\begin{figure}
\begin{tabular}{lr}
\includegraphics[width=0.5\textwidth]{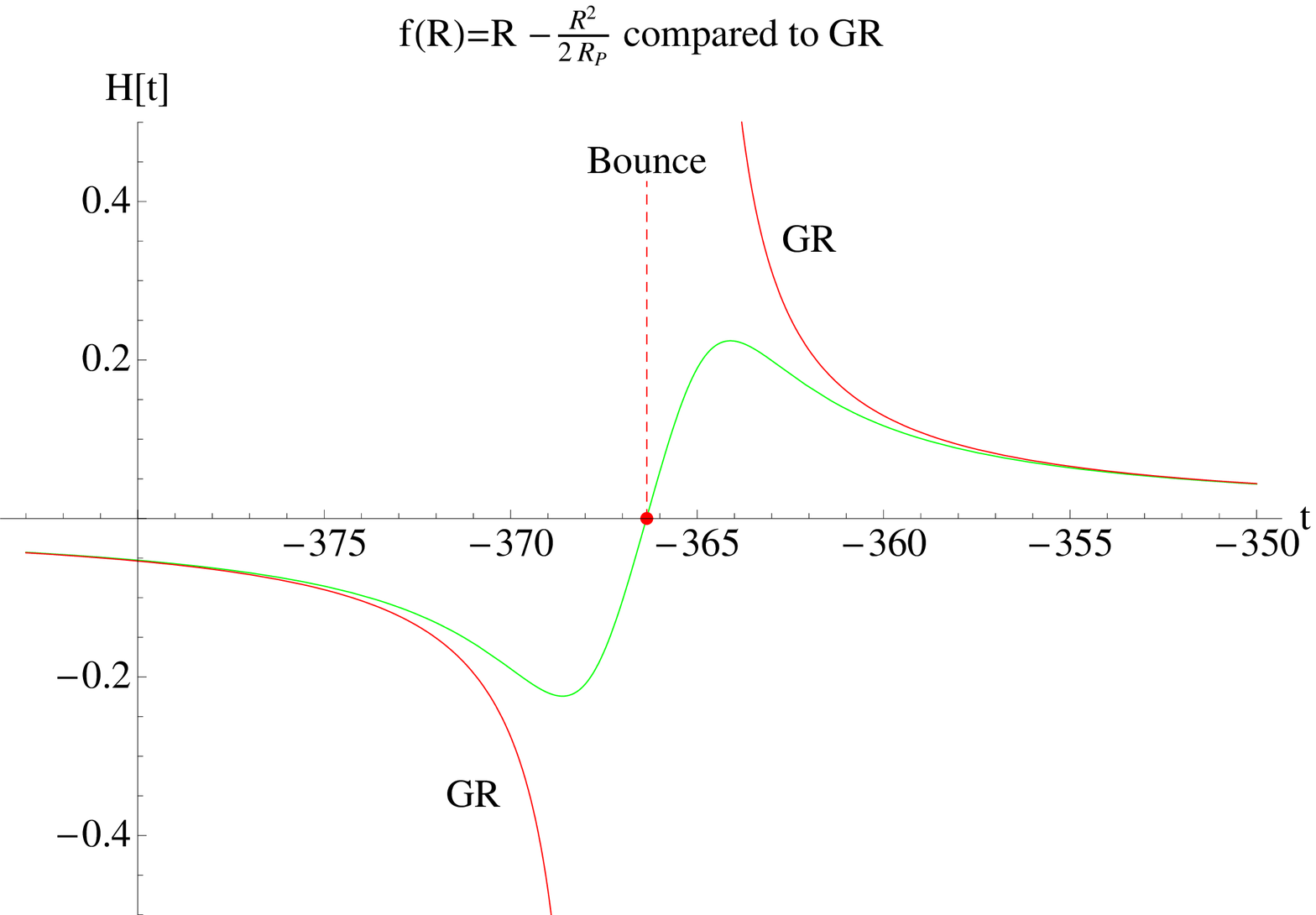} & \includegraphics[width=0.5\textwidth]{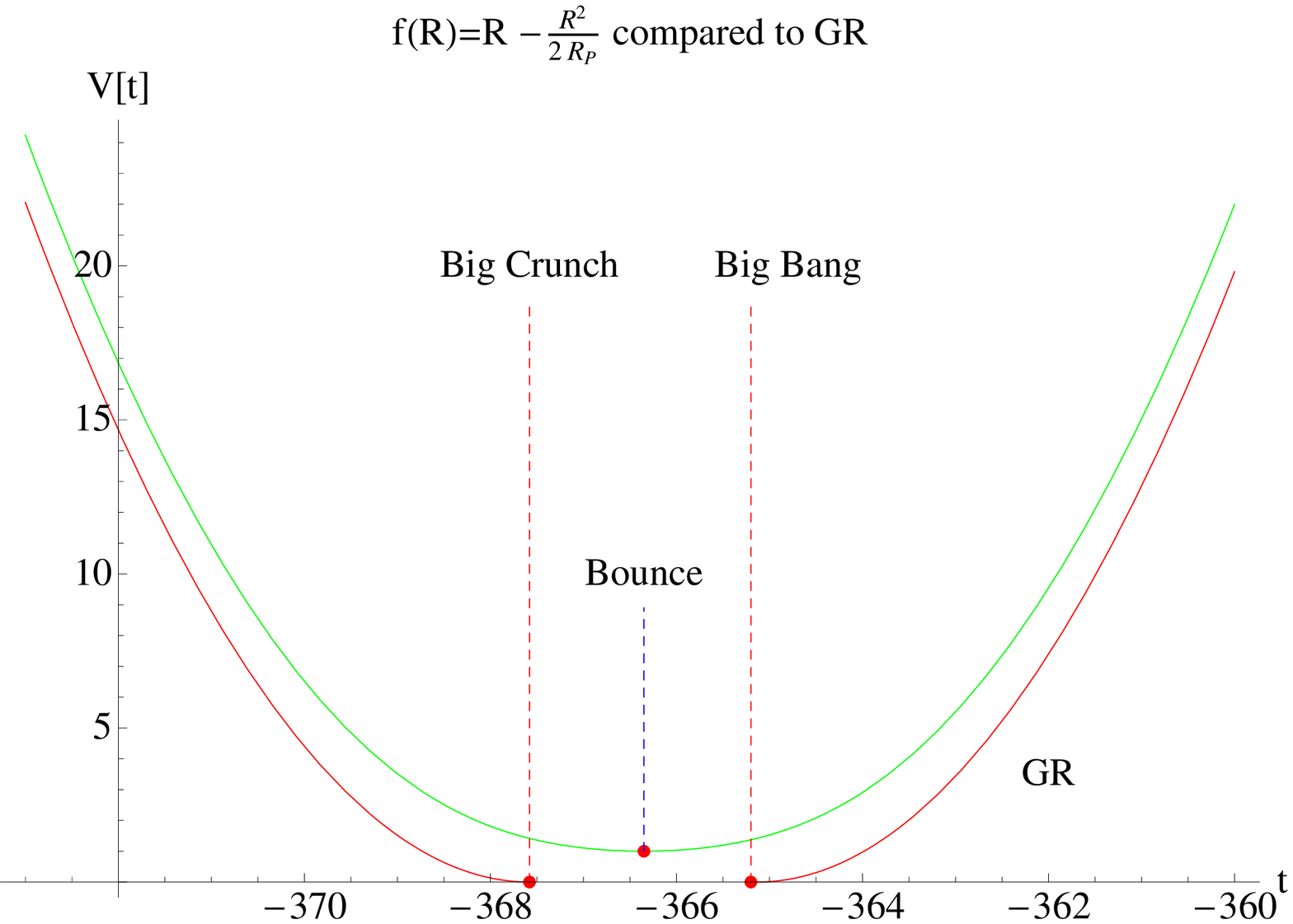}
\end{tabular}
\caption{Representation of the Hubble function (left) and volume of the Universe (right) as a function of time for the model $f(R)=R-R^2/2R_P$ in a universe filled with dust and radiation (for the numerical integration $\rho_{d,0}=10^{3}\rho_{r,0}$, and $V=10^5 V_0$). The GR solutions corresponding to a contracting branch, which ends in a big crunch, and an expanding branch, which begins with a big bang, are represented together with the bouncing solution of the Palatini model that interpolates between those singular solutions. \label{fig:iso-f(R)}}
\end{figure}

We now present the cosmological dynamics of simple $f(R)$ models to illustrate how this family of theories modifies the standard Big Bang picture of the early universe. Consider, for instance, the model\footnote{Note that the constant $a$ could be absorbed into a redefinition of $R_P$ and, therefore, only its sign is relevant.} $f(R)=R+a R^2/R_P$, where $R_P=l_P^{-2}=c^3/\hbar G$ is the Planck curvature. From the trace equation $Rf_R-2f=\kappa^2T$ (see Sec.\ref{sec:model}), we find that this model leads to the same relation between the matter and the scalar curvature as in GR, namely, $R=-\kappa^2T$. This implies that the theory behaves as GR whenever the energy density is much smaller than the Planck density scale $\rho_P\equiv R_P/\kappa^2$.  Since by definition $\theta=\frac{1}{3}\sum_i\frac{\dot a_i}{a_i}=\frac{1}{3}\frac{d}{dt}\ln a_1 a_2 a_3=\frac{1}{3}\frac{\dot V}{V}$, where $V=V_0 a_1 a_2 a_3$ represents the volume of the universe (with $V_0=V(t_0)$),  Eq. (\ref{eq:Hubble-f(R)}) for this quadratic model with dust and radiation leads to 
\begin{equation}\label{eq:Exp-cuadratic}
\theta^2=\frac{1}{9}\left(\frac{\dot V}{V}\right)^2=\frac{\left(\rho_d+\rho_r+ \frac{a\rho_d}{2\rho_P}\right)\left(1+ \frac{2a\rho_d}{\rho_P}\right)}{3\left(1- \frac{a\rho_d}{\rho_P}\right)^2} +\frac{(C_{12}^2+C_{23}^2+C_{31}^2)}{54\left(1-\frac{a\rho_d}{\rho_P}\right)^2}\left(\frac{V_0}{V(t)}\right)^2\ ,
\end{equation} 
where $\rho_d=\rho_{d,0} \left(\frac{V_0}{V(t)}\right)$ and $\rho_r=\rho_{r,0} \left(\frac{V_0}{V(t)}\right)^{4/3}$. \\
In general, an homogeneous cosmological model experiences a bounce when the expansion $\theta$ vanishes, which  implies an extremum (a maximum or a minimum) of the volume of the Universe. If $V(t)$ vanishes at some finite time, then a big bang or big crunch singularity is found, depending on whether $\dot V>0$ or $\dot V<0$ at that time. Focusing for the moment on the isotropic case, $C_{12}^2+C_{23}^2+C_{31}^2=0$, we find that a bounce occurs if $a<0$ when $\rho_d$ reaches the value $\rho_d^B=\rho_P/(2|a|)$ [see Fig.\ref{fig:iso-f(R)})] . This value of the density implies that $f_R=1-2a \rho_d/\rho_P=0$. This condition, $f_R=0$, characterizes the location of the bounce in Palatini $f(R)$ theories with a single fluid with constant equation of state [\cite{BO2010}]. For our quadratic model, in particular, bouncing solutions exist if the dynamics allows to reach the density $\rho_B=\frac{\rho_P}{2a(3w-1)}>0$. This means that for $a>0$ fluids with $w>1/3$ avoid the initial singularity, whereas  for $a<0$ it takes $w<1/3$. The case $a=0$ naturally recovers the equations of GR. It is worth noting that a cosmic bounce may arise even for presureless matter, $w=0$, if $a<0$, which implies that exotic sources of matter-energy that violate the energy conditions are not necessary to avoid the big bang singularity in this framework. The reason for this is that at high energies  gravitation may become repulsive for matter sources with $w>-1$, whereas it is attractive at low energy-densities for those same sources. \\
Note also that the pure radiation universe, $w=1/3$, is a peculiar case because it does not produce any modified dynamics in Palatini $f(R)$  theories. On physical grounds, however, it should be noted that due to quantum effects related with the trace anomaly of the electromagnetic field, a gas of photons in a $SU(N)$ gauge theory with $N_f$ fermion flavors has an effective equation of state given by 
\begin{equation}
w_{eff}^{rad}=\frac{1}{3}-\frac{5\alpha^2}{18\pi^2}\frac{\left(N_c+\frac{5}{4}N_f\right)\left(\frac{11}{3}N_c-\frac{2}{3}N_f\right)}{2+\frac{7}{2}\frac{N_c N_f}{N_c^2-1}} \ ,
\end{equation}   
where $N_c$ is the color number of the gauge theory (which has $N_c(N_c-1)$ generators) [\cite{wradeff}], [\cite{Gaetano}]. Therefore, a universe filled with photons should be able to avoid the singularity if $a>0$. \\
In physically realistic scenarios, one should consider the co-existence of several fluids and take into account the time dependence of the number of effective degrees of freedom and the transfer of energy among different species [\cite{Inflation_books4}], which leads to the possibility of having different effective fluids at different stages of the cosmic expansion. In this sense, the "dust plus radiation" model represented by (\ref{eq:Exp-cuadratic}) needs not be accurate at all times because dust particles may become relativistic at high energies and contribute to $\rho_r$ rather than to $\rho_d$. This suggests that the choice/determination of the sign of the parameter $a$ is not a trivial issue and would require a very careful and elaborate analysis (which goes beyond the scope of this introductory work). \\

\begin{figure}
\begin{tabular}{lr}
\includegraphics[width=0.5\textwidth]{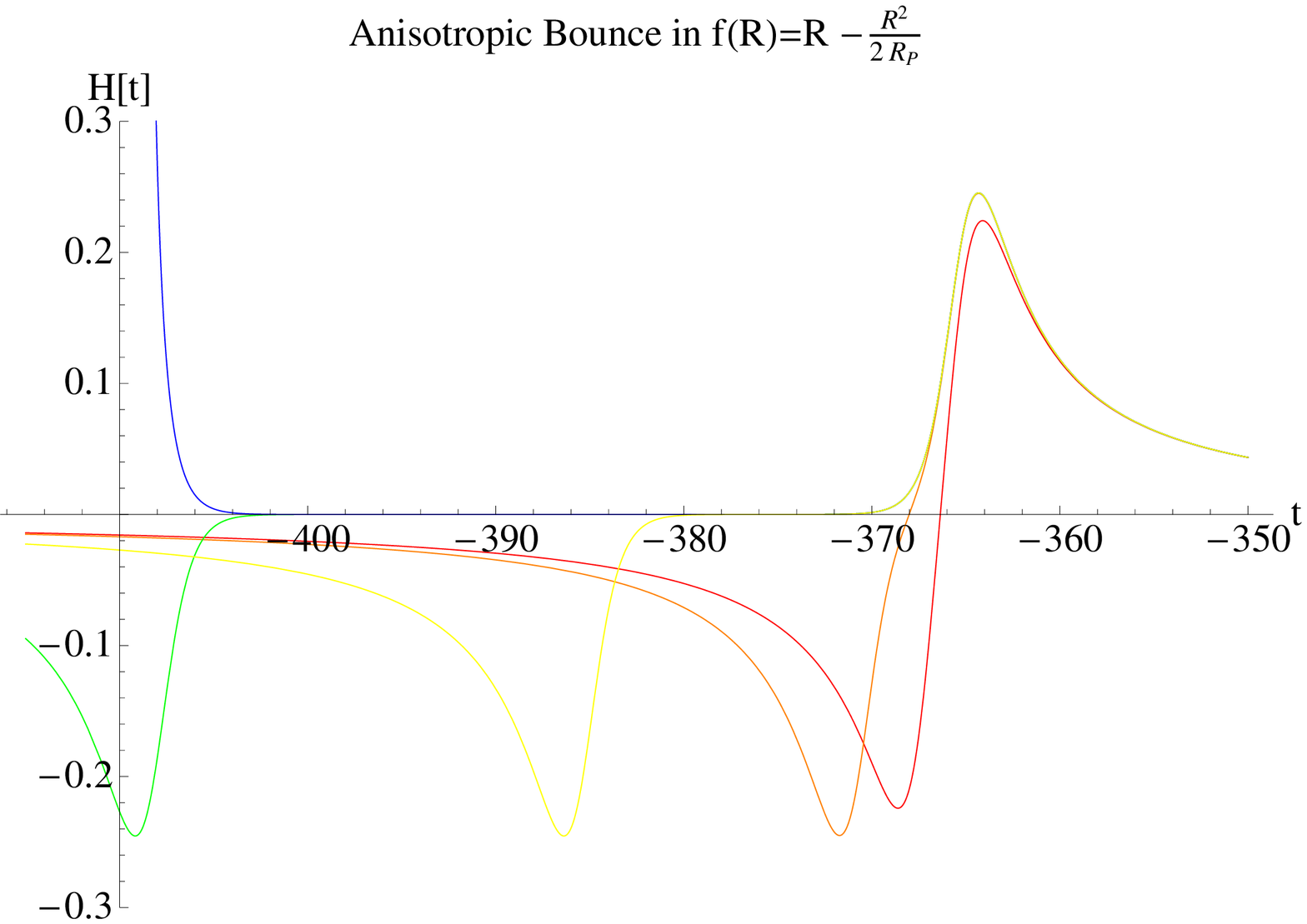} & \includegraphics[width=0.5\textwidth]{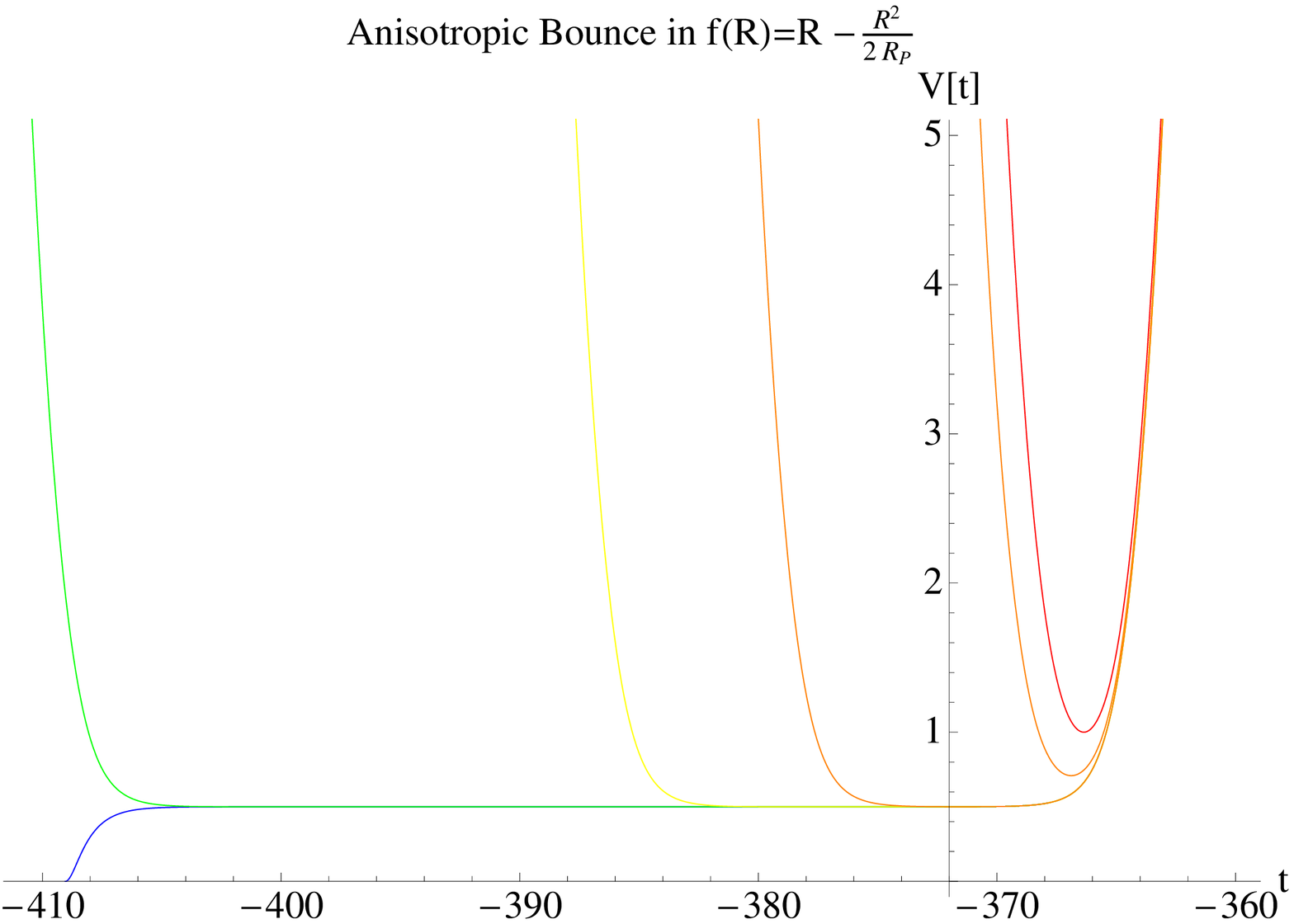}
\end{tabular}
\caption{Representation of the expansion (left) and volume of the Universe (right) as a function of time for the model $f(R)=R-R^2/2R_P$ in a universe filled with dust and radiation with anisotropies (for the numerical integration $\rho_{d,0}=10^{3}\rho_{r,0}$, and $V=10^5 V_0$). From right to left, we have plotted the bouncing cases $C^2=0, 40, 40.60211073, 40.60211073942454489657$, and the collapsing case with $C^2=40.60211073942454489658$. Fine tunning the value of $C^2$ even more should allow to keep the universe in its minimum for longer periods of time in the past, which eventually should lead to an asymptotically static solution. \label{fig:anis-f(R)}}
\end{figure} 

When anisotropies are taken into account, one finds that bouncing solutions are still possible as long as the amount of anisotropy is not too large. In Fig.\ref{fig:anis-f(R)}, we see that increasing the value of $C^2\equiv C_{12}^2+C_{23}^2+C_{31}^2$ from zero, the volume of the universe presents a minimum as long as $C^2<C^2_{c}$. If $C^2>C^2_c$, the collapse is unavoidable and $V\to 0$ in a finite time. The critical case $C^2\to C_c^2$ represents a configuration that is neither a bouncing universe nor a big bang. It corresponds to a state in which the volume of the universe remains constant in the past and expands in the future. Though this solution is clearly unstable and fine-tuned, its existence puts forward the possibility of obtaining static regular solutions corresponding to ultracompact objects, which could shed new light on the internal structure of black holes and/or topological deffects when Planck scale corrections to the gravitational action are taken into account.  It should be noted, however, that in order to obtain this asymptotically static solution one must cross from the domain where $f_R>0$ to the region where $f_R<0$. Since the shear, as defined in (\ref{eq:shear-f(R)}) for $f(R)$ theories with perfect fluids, is proportional to $1/f_R^2$, the crossing through $f_R=0$ implies a divergence in some curvature scalars of the theory. Whether this divergence is a true (or strong) physical singularity in the sense defined in [\cite{TipClarKro1}], [\cite{TipClarKro2}], [\cite{TipClarKro3}] is an open question that will be explored elsewhere.   In any case, we remark that the existence of that divergence does not have any effect on the time evolution of the expansion $\theta$, as can be seen in Fig.\ref{fig:anis-f(R)}.  \\

\subsection{Nonsingular Universes in $f(R,Q)$ Palatini theories \label{sec:f(R,Q)}}

In the previous section we have seen that Palatini $f(R)$ models are able to avoid the big bang singularity in idealized homogeneous and isotropic scenarios but run into trouble when anisotropies are present. The divergence of the shear is a generic problem for those $f(R)$ theories in which the function $f_R$ vanishes at some point, regardless of the number and equation of state of the fluids involved. Though the nature of this divergence has not been identified yet  with that of a strong singularity, which besides the divergence of some components of the Riemann, Ricci, and Weyl tensors also requires the divergence of some of their integrals, its very presence is a disturbing aspect that one would like to overcome within the framework of Palatini theories. In this sense, a natural step is to study the behavior in anisotropic scenarios of some simple generalization of the $f(R)$ family to see if the situation improves. Using Lagrangians of the form presented in  (\ref{sec:model}), we will show next that completely regular bouncing solutions exist for both isotropic and anisotropic homogeneous cosmologies. 

\subsubsection{Isotropic Universe}
Consider Eq.(\ref{eq:Hubble-iso})  particularized to the following $f(R,Q)$ Lagrangian 
\begin{equation}\label{eq:f(R,Q)}
f(R,Q)=R+a\frac{R^2}{R_P}+b\frac{Q}{R_P}
\end{equation}
For this theory, we find that $R=\kappa^2(\rho-3P)$ and $Q=Q(\rho,P)$ is given by (\ref{eq:Q}) with $\alpha\equiv b/R_P$. From now on we assume that the parameter $b$ of the Lagrangian is positive and has been absorbed into a redefinition of $R_P$, which is assumed positive. This restriction is necessary (though not sufficient) if one wants the scalar $Q$ to be bounded from above when fluids with  $w>-1$ are considered. Stated differently, when $b/R_P>0$, positivity of the square root of Eq.(\ref{eq:Q}) establishes that there may exist a maximum for the combination $\rho+P$. \\ 

\begin{figure}

\begin{center}
\begin{minipage}{0.65\textwidth}
\includegraphics[width=1\textwidth]{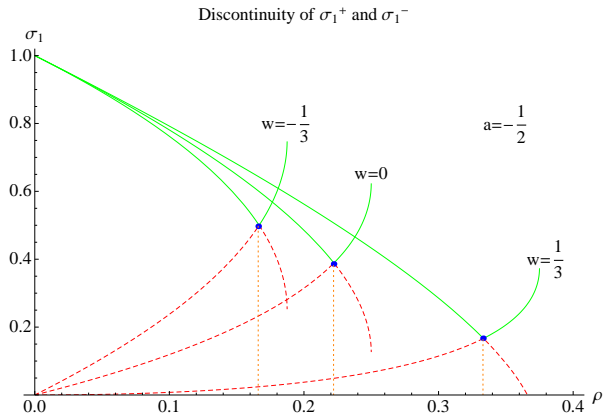}
\caption{Illustration of the need to combine the two branches of $\sigma_1$ to obtain a continuous and differentiable curve. The branch that starts at $\sigma_1=1$ has the plus sign in front of the square root (continuous green line). When the square root vanishes (at the blue dot), the function must be continued through the dashed red branch, which corresponds to the negative sign in front of the square root.  \label{fig:Sigmapm}}
\end{minipage}
\end{center}

\end{figure}

\indent In order to have (\ref{eq:Hubble-iso})  well defined, one must make sure that the choice of sign in front of the square root of $\sigma_1$ in (\ref{eq:sigmas}) is the correct one. In this sense, we find that to recover the $f(R)$ limit and GR at low curvatures, we must take the positive sign, i.e., $\sigma_1=\sigma_1^+$. However, when considering particular models, which are characterized by the constant $a$ and, for instance, a constant equation of state $w$, one realizes that the square root may reach a zero at some high density. Beyond that point, we may need to switch from $\sigma_1^+$ to $\sigma_1^-$ to guarantee that $\sigma_1$ is a continuous and differentiable function  (see Fig.\ref{fig:Sigmapm} for an illustration of this point). Bearing in mind this technical subtlety, one can then proceed to represent the Hubble function for different choices of parameters and fluid combinations to determine whether bouncing solutions exist or not. \\

The classification of the bouncing solutions of the model (\ref{eq:f(R,Q)}) with a fluid with constant $w$ was carried out in [\cite{BO2010}]. It was found that  for every value of the parameter $a$ there exist an infinite number of bouncing solutions, which depend on the particular equation of state $w$. The bouncing solutions can be divided into two large classes:
\begin{itemize}
\item {\bf Class I: $a\geq 0$.} The bounce occurs when the scalar $Q$ reaches its maximum value and happens for all equations of state satisfying the condition
\begin{equation}\label{eq:w-wmin}
w>w_{min}=\frac{a}{2+3a} \ .
\end{equation}
 From this equation it follows that a radiation dominated universe, with $w=1/3$, always bounces for any $a>0$.\\

\item {\bf Class II: $a\leq 0$.} This case is more involved because the bounce can occur either at the point where $Q$ reaches its maximum or when $\sigma_1$ vanishes. This last case can only happen at high curvatures when we are in the branch defined by  $\sigma_1=\sigma_1^-$.  To proceed with the classification, we divide this sector into several intervals:
\begin{itemize}
\item {\bf If $-1/4< a\leq 0$.} The bounce occurs if  
\begin{equation}
-\frac{1}{3}+\frac{1}{3}\sqrt{\frac{1+4a}{1+a}}<w<\infty
\end{equation}
We see that when $a=0$ we find agreement with the discussion of case I. As $a$ approaches the limiting value $-1/4$, the bouncing solutions extend up to $w\to -1/3$. 

\item {\bf If $-1/3\leq a\leq-1/4$.} Numerically one finds that the bouncing solutions cannot be extended below $w<-1$ and  occur if $-1<w<\infty$, where $w=-1$ is excluded. 

\item {\bf If $-1\leq a\leq-1/3$.} In this case, one finds numerically that the bouncing solutions are restricted to the interval $-1<w< \frac{\alpha+\beta a}{(1+3a)^2}>1$, where $\alpha=1.1335$ and $\beta=-3.3608$. 

\item {\bf If $a\leq-1$.} Similarly as the family $a\geq 0$, this set of models also allows for a simple characterization of the bouncing solutions, which correspond to the interval $-1<w<a/(2+3a)$. In the limiting case $a=-1$ we obtain the condition $-1<w<1$ (compare this with the numerical fit above, which gives $-1<w<1.12$). 

\end{itemize}

\end{itemize}

\subsection{Anisotropic Universe}\label{sec:VB}

Using Eqs. (\ref{eq:Hubble-pfluids}) and (\ref{eq:Hubble-iso}), the expansion can be written as follows:
\begin{equation}\label{eq:Exp_H}
\theta^2=H^2+\frac{1}{6}\frac{\sigma^2}{(1+\frac{3}{2}\Delta_1)^2} \ ,
\end{equation}
where $H$ represents the Hubble function in the $K=0$ isotropic case. To better understand the behavior of $\theta^2$, let us consider when and why $H^2$ vanishes. Using the results of [\cite{BO2010}] summarized above,  one finds that $H^2$ vanishes either when the density reaches the value $\rho_{Q_{max}}$ or when the function $\sigma_1$ vanishes. These two conditions imply a divergence in the quantity $(1+\frac{3}{2}\Delta_1)^2$, which appears in the denominator of $H^2$ and, therefore, force the vanishing of $H^2$ (isotropic bounce). Technically, these two types of divergences can be easily characterized. From the definition of $\Delta_1$ in (\ref{eq:D1}), one can see that $\Delta_1\sim \partial_\rho\Omega/\Omega$. Since $\Omega\equiv \sqrt{\sigma_1\sigma_2}$, it is clear that $\Delta_1$ diverges when $\sigma_1=0$. The divergence due to reaching $\rho_{Q_{max}}$ is a bit more elaborate. One must note that $\partial_\rho\Omega$ contain terms that are finite plus a term of the form $\partial_\rho\lambda$, with $\lambda$ defined below Eq. (\ref{eq:met-varRQ3}). In this $\lambda$ there is a $Q$ term hidden in the function $f(R,Q)$, which implies that $\partial_\rho\lambda\sim \partial_\rho Q/R_P$ plus other finite terms. From the definition of $Q$ it follows that $\partial_\rho Q$ has finite contributions plus the term $\partial_\rho \Phi /\sqrt{\Phi}$, where $\Phi\equiv(1+(1+2a)R/R_P)^2-4\kappa^2(\rho+P)/R_P$, which diverges when $\Phi$ vanishes. This divergence of $\partial_\rho Q$ indicates that $Q$ cannot be extended beyond the maximum value $Q_{max}$. \\
\indent Now, since the shear goes like $\sigma^2\sim 1/(\sigma_1)^2$ [see Eq.(\ref{eq:shear})], we see that the condition $\sigma_1=0$ implies a divergence on $\sigma^2$ (though $\theta^2$ remains finite). This is exactly the same type of divergence that we already found in the $f(R)$ models, where $\sigma_1\to f_R$.  Since in the $f(R)$ models the bounce can only occur when $f_R=0$, there is no way to avoid the divergence of the shear in the anisotropic case within the $f(R)$ setting. On the contrary, since the quadratic $f(R,Q)$ model (\ref{eq:f(R,Q)}) allows for a second mechanism for the bounce, which takes place at $\rho_{Q_{max}}$, there is a natural way out of the problem with the shear.  \\
	Summarizing, we conclude that for universes governed by the Lagrangian (\ref{eq:f(R,Q)}) and containing a single stiff fluid there exist completely regular bouncing solutions in the anisotropic case for $w>\frac{a}{2+3a}$ if $a\geq 0$, for $w_0<w<\infty$ if $-1/3\leq a\leq 0$, for $w_0<w<(\alpha+\beta a)/(1+3a)^2$ if $-1\leq a\leq-1/3$, and for $-1/3<w<a/(2+3a)$ if $a\leq-1$, where $w_0<0$ is defined as the equation of state for which the (isotropic) bounce occurs when $Q=Q_{max}$ and $\sigma_1=0$ simultaneously (see \cite{BO2010} for details). 
 These results imply that for $a<0$ the interval $0\leq w\leq 1/3$ is always included in the family of  completely regular isotropic and anisotropic bouncing solutions, which contain the dust and radiation cases. For $a\geq 0$, the radiation case is always nonsingular too. 

\begin{figure}
\begin{tabular}{lr}
 \includegraphics[width=0.5\textwidth]{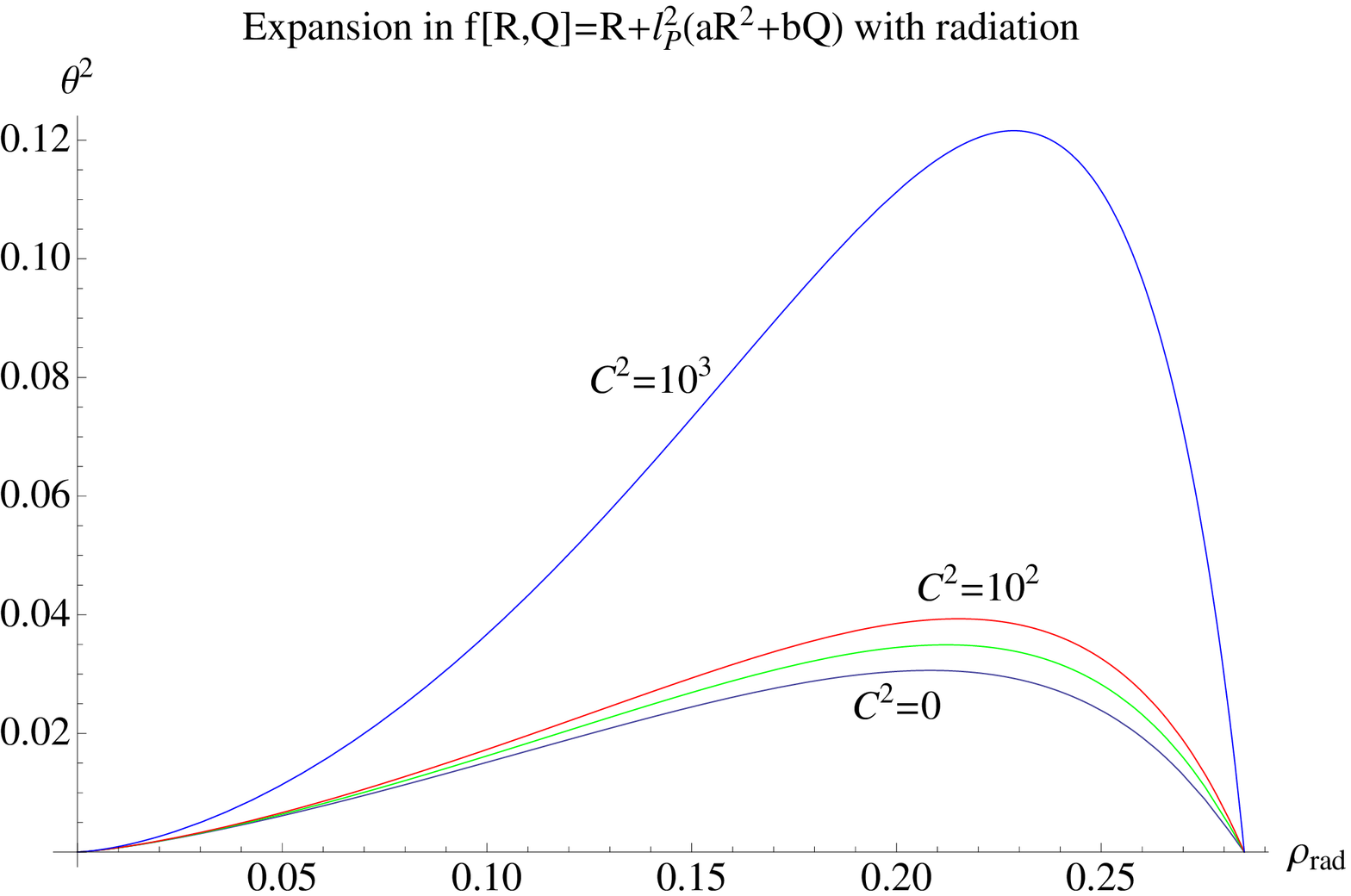} & \includegraphics[width=0.5\textwidth]{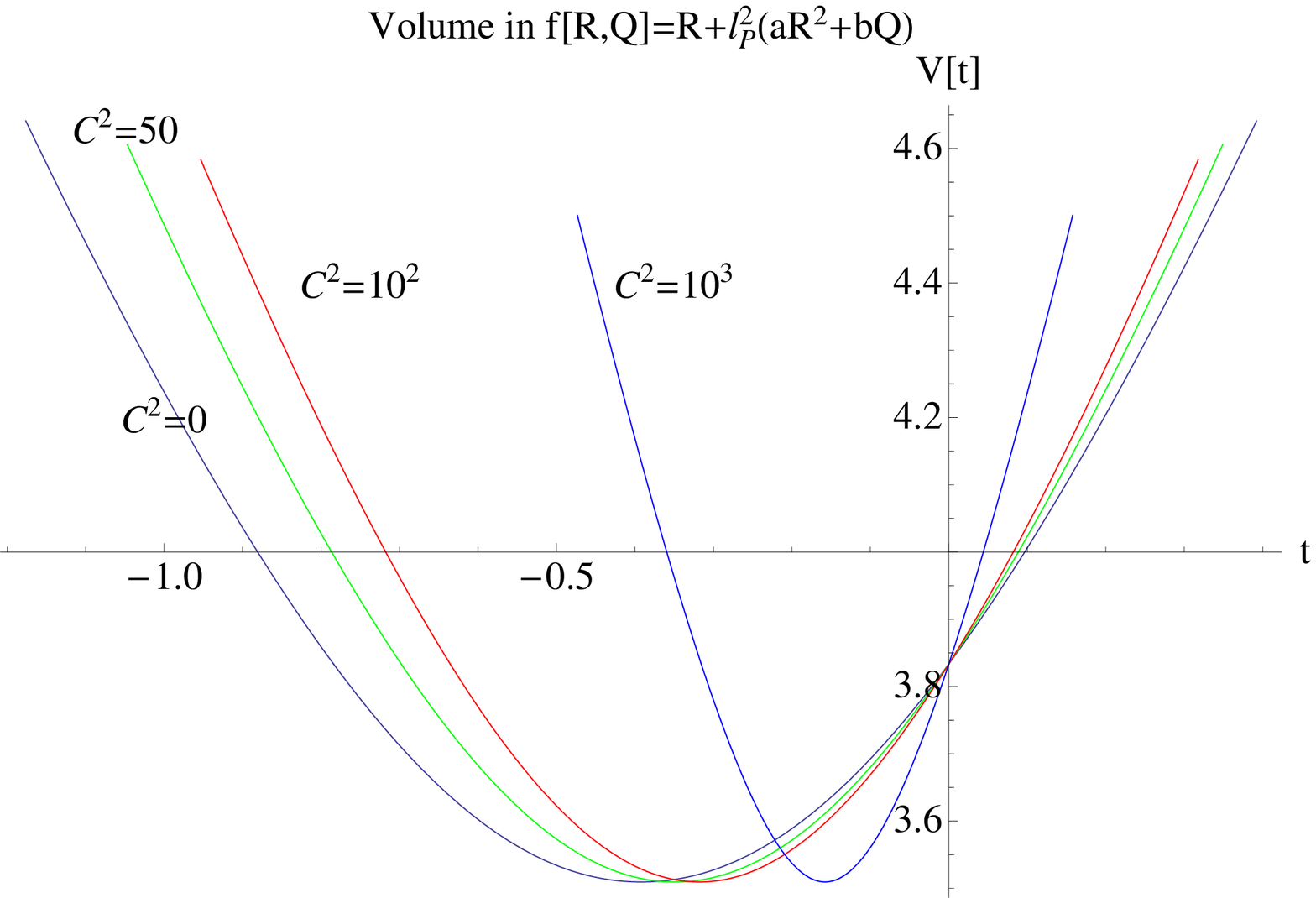}
\end{tabular}
\caption{Representation of the expansion squared (left) and volume of the Universe (right) as a function of time for the model $f(R,Q)=R-l_P^2R^2/2+l_P^2Q$ in a radiation universe  with anisotropies. We have plotted the bouncing cases $C^2=0, 50, 10^2, 10^3$. Note that the bounce always occurs at the same maximum density (minimum volume). Note also that the time spent in the bouncing region decreases as the anisotropy grows. The starting point of the time integration is chosen such that at $t=0$ the two branches of $\sigma_1$ coincide.  \label{fig:RadAnis}}
\end{figure} 
\subsection{Example: radiation universe.}

As an illustrative example, we consider here the particular case of a universe filled with radiation. Besides its obvious physical interest, this case leads to a number of algebraic simplifications that make more transparent the form of some basic definitions
\begin{eqnarray}
Q&=& \frac{3R_P^2}{8}\left[1-\frac{8\kappa^2\rho}{3R_P}-\sqrt{1-\frac{16\kappa^2\rho}{3R_P}}\right] \label{eq:Q-rad}\\
\sigma_1^{\pm}&=&\frac{1}{2}\pm \frac{1}{2\sqrt{2}} \sqrt{5 -3 \sqrt{1- \frac{16\kappa ^2\rho }{3R_P}}-\frac{24\kappa ^2\rho }{R_P}}\\
\sigma_2&=&\frac{1}{2}+\frac{1}{2\sqrt{2}} \sqrt{5 -3 \sqrt{1-\frac{16\kappa ^2\rho }{3R_P}}-\frac{8\kappa ^2\rho }{3R_P}} \\
\end{eqnarray}
Note that the coincidence of the two branches of $\sigma_1$ occurs at $\kappa^2\rho=R_P/6$, where $\sigma_1^{\pm}=\frac{1}{2}$.  It is easy to see that at low densities (\ref{eq:Q-rad}) leads to $Q\approx 4(\kappa^2\rho)^2/3+32(\kappa^2\rho)^3/9R_P+320(\kappa^2\rho)^4/27R_P^2+\ldots$, which recovers the expected result for GR, namely, $Q=3P^2+\rho^2$. From this formula we also see that the maximum value of $Q$ occurs at $\kappa^2\rho_{max}=3R_P/16$ and leads to $Q_{max}=3R_P^2/16$. At this point the shear also takes its maximum allowed value, namely, $\sigma^2_{max}=\sqrt{3/16}R_P^{3/2}(C_{12}^2+C_{23}^2+C_{31}^2)$, which is always finite. At $\rho_{max}$ the expansion vanishes producing a cosmic bounce regardless of the amount of anisotropy [see Fig.\ref{fig:RadAnis}].

\begin{figure}

\begin{center}
\begin{minipage}{0.65\textwidth}
\includegraphics[width=1\textwidth]{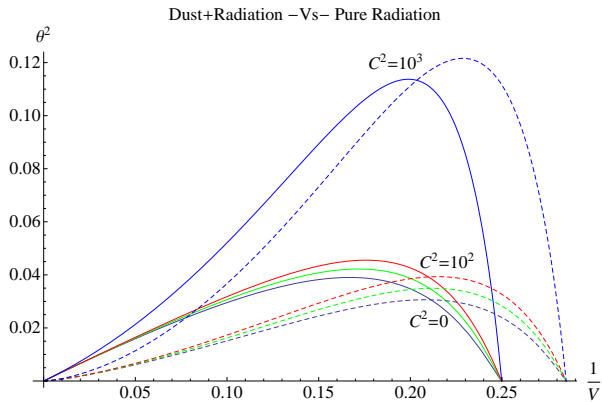}
\caption{Comparison of the expansion in a universe filled with dust and radiation ($\rho_{0,rad}=10^{-3}\rho_{0,dust}$) and a radiation dominated universe (dashed lines) for several values of the anisotropy.  \label{fig:DRandRad}}
\end{minipage}
\end{center}

\end{figure}

\section{Conclusions and open questions}

In this chapter we have tried to convey the idea that in the construction of  extended theories of gravity, one should bear in mind the fact that metric and connection are equally fundamental and independent objects. This observation allows to broaden the spectrum of available possibilities to go beyond the {\it standard model of gravitation}. In fact, any theory of gravity based on a geometry in which the connection has been forced to be given by the Christoffel symbols of the metric admits an alternative formulation in which the form of the connection is dictated by the theory itself, i.e., it is not given by convention or selected on practical grounds. \\

In our exploration of Palatini theories, we have seen that assuming that metric and connection are independent geometrical objects has non-trivial effects on the resulting field equations as compared with the usual {\it metric} formulation of the same theories. For the particular family of $f(R,Q)$ models studied here, we have seen that 
the metric is governed by second-order equations that boil down to GR in vacuum. This is in sharp contrast with the usual {\it metric} formulation of those same theories, where one finds fourth-order derivatives of the metric (see, for instance, [\cite{Anderson}] for a detailed analysis of the cosmology of the quadratic model (\ref{eq:f(R,Q)}) in metric formalism). The absence of higher-order derivatives in the Palatini formulation is a remarkable point that seems not to have been sufficiently appreciated in the literature. In fact,  having second-order field equations is very important because it automatically implies the absence of ghosts and other dynamical instabilities. In this sense, it should be noted that Lovelock\footnote{We would like to mention that when Lovelock theories are formulated \`{a} la Palatini, the resulting field equations are exactly the same as one finds in their usual {\it metric} formulation [\cite{MBG}]. } theories [\cite{Zanelli}],  which are generally regarded as the natural extension of the Einstein-Hilbert Lagrangian to higher dimensions, have received a lot of attention in the literature because they are seen as the most general actions for gravity that give at most second-order field equations for the metric. As we have seen here, this property is shared (at least) by all Palatini theories of the $f(R,Q)$ type (with or without torsion). This puts forward that Palatini theories, are natural candidates to explore new dynamics beyond GR. \\

Before concluding, we would like to stress the fact that the quadratic Palatini model (\ref{eq:f(R,Q)}) is able to avoid the big bang singularity in very natural situations, such as in pure radiation, pure dust, or dust plus radiation universes with or without anisotropies (see Fig.\ref{fig:DRandRad}). This observation has been possible thanks to the formulas presented in section \ref{sec:cosmo}, where we have extended the analysis carried out in [\cite{BO2010}] for a single perfect fluid with constant equation of state to include several perfect fluids with arbitrary equation of state $w(\rho)$. This allows to explore the dynamics of realistic cosmological models with several fluids  and is a necessary step prior to the consideration of the growth and evolution of inhomogeneities in these nonsingular backgrounds. \\
Though the model (\ref{eq:f(R,Q)}) has been proposed on grounds of mathematical simplicity and motivated by the form of the effective action provided by perturbative quantization schemes in curved backgrounds, its ability to successfully deal with cosmological [\cite{BO2010}] and black hole singularities [\cite{Olmo:2011aw}] as well as other aspects of quantum gravity phenomenology [\cite{Olmo:2011sw}]   demands further theoretical work  to provide a more solid ground to it. In this sense, we note that the effective dynamics of loop quantum cosmology [\cite{lqc}] in a Friedmann-Robertson-Walker background filled with a massless scalar can be exactly reproduced by a Palatini $f(R)$ theory [\cite{Olmo:2008nf}]. The extension of that result to more general space-times and matter sources could shed new light on the potential relation of  (\ref{eq:f(R,Q)}) with a more fundamental theory of quantum gravity. All these open questions will be considered in detail elsewhere.\\

This work has been supported by the Spanish grants FIS2008-06078-C03-02, FIS2011-29813-C02-02, the Consolider Program CPAN (CSD2007-00042), and the JAE-doc program of the Spanish Research Council (CSIC).

\end{document}